\newif\ifcomments  
\newif\ifanonymous 
\newif\ifarxiv     

\commentsfalse
\anonymousfalse
\arxivtrue

\ifarxiv\pdfoutput=1\fi
\documentclass[
  a4paper,UKenglish,cleverref,
  \ifanonymous anonymous,\fi
]{lipics-v2021}

\usepackage{enumitem,multicol,xspace,doi}
\usepackage{mathpartir,mathtools,stmaryrd}
\usepackage[capitalize,noabbrev,nameinlink]{cleveref}
\usepackage[flushmargin,multiple]{footmisc} 

\ifarxiv
\usepackage[frozencache,cachedir=.]{minted}
\else
\usepackage{minted}
\fi

\usepackage{bm}

\usepackage[supertabular]{ottalt}
\usepackage{draft}

\ifarxiv
\newcommand{\monoscale}{0.85}
\usepackage[scale=\monoscale]{noto-mono}
\DeclareUnicodeCharacter{0394}{\scalebox{\monoscale}{$\Delta$}}
\DeclareUnicodeCharacter{03A9}{\scalebox{\monoscale}{$\Omega$}}
\DeclareUnicodeCharacter{03BB}{\scalebox{\monoscale}{$\lambda$}}
\DeclareUnicodeCharacter{03C3}{\scalebox{\monoscale}{$\sigma$}}
\DeclareUnicodeCharacter{03C4}{\scalebox{\monoscale}{$\tau$}}
\DeclareUnicodeCharacter{2081}{$_{1}$}
\DeclareUnicodeCharacter{2082}{$_{2}$}
\DeclareUnicodeCharacter{2083}{$_{3}$}
\DeclareUnicodeCharacter{2113}{\scalebox{\monoscale}{$\ell$}}
\DeclareUnicodeCharacter{2192}{\scalebox{\monoscale}[1]{$\shortrightarrow$}}
\DeclareUnicodeCharacter{21D2}{\scalebox{\monoscale}{$\Rightarrow$}}
\DeclareUnicodeCharacter{21D4}{\scalebox{\monoscale}{$\Leftrightarrow$}}
\DeclareUnicodeCharacter{2200}{\scalebox{\monoscale}{$\forall$}}
\DeclareUnicodeCharacter{2248}{\scalebox{\monoscale}{$\approx$}}
\DeclareUnicodeCharacter{2294}{\scalebox{\monoscale}{$\sqcup$}}
\DeclareUnicodeCharacter{22A5}{\scalebox{\monoscale}{$\bot$}}
\DeclareUnicodeCharacter{22C6}{\scalebox{\monoscale}{$\star$}}
\DeclareUnicodeCharacter{1D7D8}{\scalebox{\monoscale}{$\mathbf{0}$}}
\DeclareUnicodeCharacter{1D7D9}{\scalebox{\monoscale}{$\mathbf{1}$}}
\DeclareUnicodeCharacter{1D7DA}{\scalebox{\monoscale}{$\mathbf{2}$}}
\DeclareUnicodeCharacter{1D7DB}{\scalebox{\monoscale}{$\mathbf{3}$}}
\else
\usepackage{fontspec}
\fi

\hypersetup{
  colorlinks=true,
  urlcolor=blue,
  citecolor=violet,
  backref=true
}

\DeclarePairedDelimiter{\interp}{\llbracket}{\rrbracket}

\newcommand{\citep}[1]{\cite{#1}}
\newcommand{\citet}[1]{\cite{#1}}
\newcommand{\publicrepo}{\url{https://github.com/plclub/StraTT}}
\newcommand{\lang}{\textsf{StraTT}\@\xspace}
\newcommand{\sublang}{\textsf{subStraTT}\@\xspace}
\newcommand{\langue}{Stratified Type Theory\@\xspace}
\newcommand{\ie}{\textit{i.e.}\@\xspace}

\newcommand{\welltyped}{well-\hspace{0pt}typed\@\xspace}
\newcommand{\wellfounded}{well-\hspace{0pt}founded\@\xspace}
\newcommand{\wellfoundedness}{well-\hspace{0pt}foundedness\@\xspace}
\newcommand{\wellformedness}{well-\hspace{0pt}formedness\@\xspace}

\newcommand{\crude}{crude-\hspace{0pt}but-\hspace{0pt}effective\@\xspace}
\newcommand{\type}{\mathrel{\mathsf{\bf type}}}
\newcommand{\footfile}[1]{%
  \raggedright
  \ifanonymous
    \texttt{#1}
  \else
    \href{https://github.com/plclub/StraTT/tree/main/#1}{\texttt{#1}}
  \fi
}
\newcommand{\footfilethm}[2]{%
  \raggedright
  \ifanonymous
    \texttt{#1:#2}
  \else
    \href{https://github.com/plclub/StraTT/tree/main/#1}{\texttt{#1}}\texttt{:#2}
  \fi
}

\newlength{\punctwidth}
\newcommand{\punctstack}[1]{#1%
  \settowidth{\punctwidth}{#1}%
  \hspace*{-\the\punctwidth}%
}
\newcommand{\stackpunct}[2]{%
  \settowidth{\punctwidth}{$^{#1}$}%
  \hspace*{-\the\punctwidth}%
  #2%
}
\newcommand{\stack}[1]{%
  \settowidth{\punctwidth}{#1}%
  \hspace*{-\the\punctwidth}%
}

\newtheorem{axiom}[theorem]{Axiom}

\ifcomments
\newnote{scw}{blue} 
\newnote{jcxz}{teal} 
\else
\newcommand{\scw}[1]{}
\newcommand{\jcxz}[1]{}
\fi

\title{Stratified Type Theory}
\authorrunning{J. Chan, S. Weirich}
\Copyright{Jonathan Chan, Stephanie Weirich}
\hideLIPIcs
\nolinenumbers

\author{Jonathan Chan}
  {University of Pennsylvania, Philadelphia, USA}
  {jcxz@seas.upenn.edu}
  {0000-0003-0830-3180}
  {}

\author{Stephanie Weirich}
  {University of Pennsylvania, Philadelphia, USA}
  {sweirich@seas.upenn.edu}
  {0000-0002-6756-9168}
  {}

\supplementdetails[subcategory={source code},
  swhid={swh:1:dir:0db8321cc2e274e561ca0f7dc78e9eae406b4e74} 
]{Software}{https://github.com/plclub/StraTT}

\begin{document}

\maketitle

\begin{abstract}
A hierarchy of type universes is a rudimentary ingredient in
the type theories of many proof assistants to prevent the logical inconsistency
resulting from combining dependent functions and the type-in-type rule.
In this work, we argue that a universe hierarchy is not the \emph{only} option
for a type theory with a type universe.
Taking inspiration from Leivant's Stratified System F,
we introduce \textbf{\langue} (\lang),
where rather than stratifying universes by levels,
we stratify typing judgements and restrict the domain of dependent functions
to strictly lower levels.
Even with type-in-type, this restriction suffices to enforce consistency.

In \lang, we consider a number of extensions beyond just stratified dependent functions.
First, the subsystem \sublang employs McBride's \crude stratification
(also known as displacement) as a simple form of level polymorphism
where global definitions with concrete levels can be displaced uniformly to any higher level.
Second, to recover some expressivity lost due to the restriction on dependent function domains,
the full \lang includes a separate nondependent function type with a \emph{floating} domain
whose level matches that of the overall function type.
Finally, we have implemented a prototype type checker for \lang
extended with datatypes and inference for level and displacement annotations,
along with a small core library.

We have proven \sublang to be consistent and \lang to be type safe,
but consistency of the full \lang remains an open problem,
largely due to the interaction between floating functions and cumulativity of judgements.
Nevertheless, we believe \lang to be consistent,
and as evidence have verified the failure of some well-known type-theoretic paradoxes
using our implementation.
\end{abstract}

\newcommand{\ottdrule}[4][]{{\displaystyle\frac{\begin{array}{l}#2\end{array}}{#3}\quad\ottdrulename{#4}}}
\newcommand{\ottusedrule}[1]{\[#1\]}
\newcommand{\ottpremise}[1]{ #1 \\}
\newenvironment{ottdefnblock}[3][]{ \framebox{\mbox{#2}} \quad #3 \\[0pt]}{}
\newenvironment{ottfundefnblock}[3][]{ \framebox{\mbox{#2}} \quad #3 \\[0pt]\begin{displaymath}\begin{array}{l}}{\end{array}\end{displaymath}}
\newcommand{\ottfunclause}[2]{ #1 \equiv #2 \\}
\newcommand{\ottnt}[1]{\mathit{#1}}
\newcommand{\ottmv}[1]{\mathit{#1}}
\newcommand{\ottkw}[1]{\mathbf{#1}}
\newcommand{\ottsym}[1]{#1}
\newcommand{\ottcom}[1]{\text{#1}}
\newcommand{\ottdrulename}[1]{\textsc{#1}}
\newcommand{\ottcomplu}[5]{\overline{#1}^{\,#2\in #3 #4 #5}}
\newcommand{\ottcompu}[3]{\overline{#1}^{\,#2<#3}}
\newcommand{\ottcomp}[2]{\overline{#1}^{\,#2}}
\newcommand{\ottgrammartabular}[1]{\begin{supertabular}{llcllllll}#1\end{supertabular}}
\newcommand{\ottmetavartabular}[1]{\begin{supertabular}{ll}#1\end{supertabular}}
\newcommand{\ottrulehead}[3]{$#1$ & & $#2$ & & & \multicolumn{2}{l}{#3}}
\newcommand{\ottprodline}[6]{& & $#1$ & $#2$ & $#3 #4$ & $#5$ & $#6$}
\newcommand{\ottfirstprodline}[6]{\ottprodline{#1}{#2}{#3}{#4}{#5}{#6}}
\newcommand{\ottlongprodline}[2]{& & $#1$ & \multicolumn{4}{l}{$#2$}}
\newcommand{\ottfirstlongprodline}[2]{\ottlongprodline{#1}{#2}}
\newcommand{\ottbindspecprodline}[6]{\ottprodline{#1}{#2}{#3}{#4}{#5}{#6}}
\newcommand{\ottprodnewline}{\\}
\newcommand{\ottinterrule}{\\[5.0mm]}
\newcommand{\ottafterlastrule}{\\}
\newcommand{\ottmetavars}{
\ottmetavartabular{
 $ \mathit{var} ,\, \mathit{x} ,\, \mathit{y} ,\, \mathit{z} $ & \ottcom{variables} \\
 $ \ottmv{n} $ &  \\
}}

\newcommand{\ottassn}{
\ottrulehead{\ottnt{assn}}{::=}{\ottcom{context assumptions}}\ottprodnewline
\ottfirstprodline{|}{ :^{ \ottnt{k} }  \ottnt{A} }{}{}{}{}}

\newcommand{\ottcontext}{
\ottrulehead{\ottnt{context}  ,\ \Gamma}{::=}{\ottcom{contexts}}\ottprodnewline
\ottfirstprodline{|}{ \mathit{x}   \ottnt{assn} }{}{}{}{\ottcom{singleton}}\ottprodnewline
\ottprodline{|}{\varnothing}{}{}{}{\ottcom{empty}}\ottprodnewline
\ottprodline{|}{ \Gamma ,  \Gamma' }{}{}{}{\ottcom{append}}}

\newcommand{\ottdef}{
\ottrulehead{\ottnt{def}}{::=}{\ottcom{signature definitions}}\ottprodnewline
\ottfirstprodline{|}{ :^{ \ottnt{k} }  \ottnt{A}  \coloneqq  \ottnt{a} }{}{}{}{}}

\newcommand{\ottsignature}{
\ottrulehead{\ottnt{signature}  ,\ \Delta}{::=}{\ottcom{signatures}}\ottprodnewline
\ottfirstprodline{|}{ \mathit{x}   \ottnt{def} }{}{}{}{\ottcom{singleton}}\ottprodnewline
\ottprodline{|}{\varnothing}{}{}{}{\ottcom{empty}}\ottprodnewline
\ottprodline{|}{ \Delta ,  \Delta' }{}{}{}{\ottcom{append}}}

\newcommand{\ottvalue}{
\ottrulehead{\ottnt{value}  ,\ \ottnt{v}}{::=}{}\ottprodnewline
\ottfirstprodline{|}{\ottnt{A} \, \ottnt{ne}}{}{}{}{}\ottprodnewline
\ottprodline{|}{\ottsym{\mbox{$\backslash{}$}}  \ottnt{cl}}{}{}{}{}\ottprodnewline
\ottprodline{|}{\Pi \, \ottnt{v} \, \ottnt{cl}}{}{}{}{}\ottprodnewline
\ottprodline{|}{\star}{}{}{}{}}

\newcommand{\ottnf}{
\ottrulehead{\ottnt{nf}}{::=}{}\ottprodnewline
\ottfirstprodline{|}{\ottnt{A} \, \ottnt{v}}{}{}{}{}}

\newcommand{\ottne}{
\ottrulehead{\ottnt{ne}}{::=}{}\ottprodnewline
\ottfirstprodline{|}{\ottsym{[}  \ottnt{i}  \ottsym{]}}{}{}{}{}\ottprodnewline
\ottprodline{|}{\ottnt{ne} \, \ottnt{nf}}{}{}{}{}}

\newcommand{\ottclos}{
\ottrulehead{\ottnt{clos}  ,\ \ottnt{cl}}{::=}{}}

\newcommand{\ottenv}{
\ottrulehead{\ottnt{env}  ,\ \ottnt{rho}}{::=}{}}

\newcommand{\ottdefonly}{
\ottrulehead{\ottnt{defonly}}{::=}{}\ottprodnewline
\ottfirstprodline{|}{ \ottnt{pat}  \coloneqq  \ottnt{b} }{}{}{}{\ottcom{signature definitions without type}}}

\newcommand{\ottnat}{
\ottrulehead{\ottnt{nat}  ,\ \ottnt{i}  ,\ \ottnt{j}  ,\ \ottnt{k}}{::=}{\ottcom{natural numbers}}\ottprodnewline
\ottfirstprodline{|}{ \ell }{}{}{}{}}

\newcommand{\ottsigma}{
\ottrulehead{\sigma}{::=}{\ottcom{simultaneous substitution}}}

\newcommand{\ottparams}{
\ottrulehead{\ottnt{params}}{::=}{}\ottprodnewline
\ottfirstprodline{|}{ {} }{}{}{}{}\ottprodnewline
\ottprodline{|}{ \ottnt{params}  \; (\mathit{ \mathit{x} }  \ottnt{assn} ) }{}{}{}{}\ottprodnewline
\ottprodline{|}{ \ottnt{params}  \; (\mathit{ \mathit{x} } :  \ottnt{A} ) }{}{}{}{}}

\newcommand{\ottdecl}{
\ottrulehead{\ottnt{decl}}{::=}{}\ottprodnewline
\ottfirstprodline{|}{ \mathsf{ \mathit{x} } :^{ \ottnt{k} }  \ottnt{A} }{}{}{}{}\ottprodnewline
\ottprodline{|}{ \mathsf{ \mathit{x} } :  \ottnt{A} }{}{}{}{}\ottprodnewline
\ottprodline{|}{ \mathsf{a_0} :^{ \ottnt{k} }  \ottnt{A} }{}{}{}{}}

\newcommand{\ottpat}{
\ottrulehead{\ottnt{pat}}{::=}{}\ottprodnewline
\ottfirstprodline{|}{ \mathsf{ \mathit{x} } }{}{}{}{}\ottprodnewline
\ottprodline{|}{ \mathsf{ \mathit{x} }^{ \ottnt{i} } }{}{}{}{}\ottprodnewline
\ottprodline{|}{ \ottnt{pat} \; \ottnt{a} }{}{}{}{}\ottprodnewline
\ottprodline{|}{ \mathsf{a_0} }{}{}{}{}}

\newcommand{\otttm}{
\ottrulehead{\ottnt{tm}  ,\ \ottnt{A}  ,\ \ottnt{B}  ,\ \ottnt{C}  ,\ \ottnt{a}  ,\ \ottnt{b}  ,\ \ottnt{c}}{::=}{\ottcom{terms and types}}\ottprodnewline
\ottfirstprodline{|}{ \star }{}{}{}{\ottcom{universe}}\ottprodnewline
\ottprodline{|}{ \ottnt{A}  \mathrel{\rightarrow}  \ottnt{B} }{}{}{}{\ottcom{nondependent function type}}\ottprodnewline
\ottprodline{|}{ \Pi \mathit{ \mathit{x} }\!:^{ \ottnt{j} }\! \ottnt{A}  \mathpunct{.}  \ottnt{B} }{}{\textsf{bind}\; \mathit{x}\; \textsf{in}\; \ottnt{B}}{}{\ottcom{dependent function type}}\ottprodnewline
\ottprodline{|}{ \lambda \mathit{ \mathit{x} } \mathpunct{.}  \ottnt{a} }{}{\textsf{bind}\; \mathit{x}\; \textsf{in}\; \ottnt{a}}{}{\ottcom{function}}\ottprodnewline
\ottprodline{|}{ \ottnt{a}  \;  \ottnt{b} }{}{}{}{\ottcom{application}}\ottprodnewline
\ottprodline{|}{ \bot }{}{}{}{\ottcom{empty type}}\ottprodnewline
\ottprodline{|}{ \mathsf{absurd}( \ottnt{b} ) }{}{}{}{\ottcom{ex falso quodlibet}}\ottprodnewline
\ottprodline{|}{ \mathit{ \mathit{x} } }{}{}{}{\ottcom{variable}}\ottprodnewline
\ottprodline{|}{ \mathit{ \mathit{x} }^{ \ottnt{i} } }{}{}{}{\ottcom{displaced constants}}\ottprodnewline
\ottprodline{|}{ \ottnt{a}  \mathopen{ \{ }  \ottnt{b}  / \mathit{ \mathit{x} } \mathclose{ \} } } {\textsf{S}}{}{}{}\ottprodnewline
\ottprodline{|}{ \ottnt{a} } {\textsf{S}}{}{}{\ottcom{parsing precedence is hard}}\ottprodnewline
\ottprodline{|}{ \mathsf{ \mathit{x} } }{}{}{}{}\ottprodnewline
\ottprodline{|}{ \mathsf{ \mathit{x} }^{ \ottnt{i} } }{}{}{}{}\ottprodnewline
\ottprodline{|}{ \ottnt{a}  =  \ottnt{b} }{}{}{}{}\ottprodnewline
\ottprodline{|}{ \textcolor{red}{ \underline{ \ottnt{tm} } } } {\textsf{S}}{}{}{}\ottprodnewline
\ottprodline{|}{ \ottnt{a} \{ \sigma \} }{}{}{}{}\ottprodnewline
\ottprodline{|}{ \sigma [  \mathit{x}  ] }{}{}{}{}\ottprodnewline
\ottprodline{|}{ \mathbf{case} \;  \ottnt{a}  \; \mathbf{of} }{}{}{}{}\ottprodnewline
\ottprodline{|}{ \ottnt{pat}  \mathrel{\Rightarrow}  \ottnt{a} }{}{}{}{}\ottprodnewline
\ottprodline{|}{ \Lambda \ell \mathpunct{.}  \ottnt{a} }{}{}{}{}\ottprodnewline
\ottprodline{|}{ {} }{}{}{}{}\ottprodnewline
\ottprodline{|}{ \mathsf{A_0} }{}{}{}{}\ottprodnewline
\ottprodline{|}{ \mathsf{a_0} }{}{}{}{}\ottprodnewline
\ottprodline{|}{ \mathsf{MkA_0} }{}{}{}{}}

\newcommand{\ottconstr}{
\ottrulehead{\ottnt{constr}}{::=}{}\ottprodnewline
\ottfirstprodline{|}{ \mathsf{ \mathit{x} } :^{ \ottnt{k} }  \ottnt{A} }{}{}{}{}\ottprodnewline
\ottprodline{|}{ \mathsf{A_0} :^{ \ottnt{k} }  \ottnt{A} }{}{}{}{}}

\newcommand{\ottdatadef}{
\ottrulehead{\ottnt{datadef}}{::=}{}\ottprodnewline
\ottfirstprodline{|}{ \mathbf{data} \; \mathsf{ \mathit{x} }  \ottnt{params}  :^{ \ottnt{k} }  \ottnt{A}  \; \mathbf{where} }{}{}{}{}\ottprodnewline
\ottprodline{|}{ \mathbf{data} \; \mathsf{A_0}  \ottnt{params}  :^{ \ottnt{k} }  \ottnt{A}  \; \mathbf{where} }{}{}{}{}}

\newcommand{\ottgrammar}{\ottgrammartabular{
\ottassn\ottinterrule
\ottcontext\ottinterrule
\ottdef\ottinterrule
\ottsignature\ottinterrule
\ottvalue\ottinterrule
\ottnf\ottinterrule
\ottne\ottinterrule
\ottclos\ottinterrule
\ottenv\ottinterrule
\ottdefonly\ottinterrule
\ottnat\ottinterrule
\ottsigma\ottinterrule
\ottparams\ottinterrule
\ottdecl\ottinterrule
\ottpat\ottinterrule
\otttm\ottinterrule
\ottconstr\ottinterrule
\ottdatadef\ottafterlastrule
}}

\newcommand{\ottdruleEXXRefl}[1]{\ottdrule[#1]{%
}{
 \ottnt{a}   \equiv   \ottnt{a} }{%
{\ottdrulename{E\_Refl}}{}%
}}

\newcommand{\ottdruleEXXSym}[1]{\ottdrule[#1]{%
\ottpremise{ \ottnt{b}   \equiv   \ottnt{a} }%
}{
 \ottnt{a}   \equiv   \ottnt{b} }{%
{\ottdrulename{E\_Sym}}{}%
}}

\newcommand{\ottdruleEXXTrans}[1]{\ottdrule[#1]{%
\ottpremise{ \ottnt{a}   \equiv   \ottnt{b} }%
\ottpremise{ \ottnt{b}   \equiv   \ottnt{c} }%
}{
 \ottnt{a}   \equiv   \ottnt{c} }{%
{\ottdrulename{E\_Trans}}{}%
}}

\newcommand{\ottdruleEXXBeta}[1]{\ottdrule[#1]{%
}{
  \ottsym{(}   \lambda \mathit{ \mathit{x} } \mathpunct{.}  \ottnt{b}   \ottsym{)}  \;  \ottnt{a}    \equiv    \ottnt{b}  \mathopen{ \{ }  \ottnt{a}  / \mathit{ \mathit{x} } \mathclose{ \} }  }{%
{\ottdrulename{E\_Beta}}{}%
}}

\newcommand{\ottdruleEXXArrow}[1]{\ottdrule[#1]{%
\ottpremise{ \ottnt{A}   \equiv   \ottnt{A'} }%
\ottpremise{ \ottnt{B}   \equiv   \ottnt{B'} }%
}{
  \ottnt{A}  \mathrel{\rightarrow}  \ottnt{B}    \equiv    \ottnt{A'}  \mathrel{\rightarrow}  \ottnt{B'}  }{%
{\ottdrulename{E\_Arrow}}{}%
}}

\newcommand{\ottdruleEXXPi}[1]{\ottdrule[#1]{%
\ottpremise{ \ottnt{A}   \equiv   \ottnt{A'} }%
\ottpremise{ \ottnt{B}   \equiv   \ottnt{B'} }%
}{
  \Pi \mathit{ \mathit{x} }\!:^{ \ottnt{k} }\! \ottnt{A}  \mathpunct{.}  \ottnt{B}    \equiv    \Pi \mathit{ \mathit{x} }\!:^{ \ottnt{k} }\! \ottnt{A'}  \mathpunct{.}  \ottnt{B'}  }{%
{\ottdrulename{E\_Pi}}{}%
}}

\newcommand{\ottdruleEXXAbs}[1]{\ottdrule[#1]{%
\ottpremise{ \ottnt{b}   \equiv   \ottnt{b'} }%
}{
  \lambda \mathit{ \mathit{x} } \mathpunct{.}  \ottnt{b}    \equiv    \lambda \mathit{ \mathit{x} } \mathpunct{.}  \ottnt{b'}  }{%
{\ottdrulename{E\_Abs}}{}%
}}

\newcommand{\ottdruleEXXApp}[1]{\ottdrule[#1]{%
\ottpremise{ \ottnt{a}   \equiv   \ottnt{a'} }%
\ottpremise{ \ottnt{b}   \equiv   \ottnt{b'} }%
}{
  \ottnt{b}  \;  \ottnt{a}    \equiv    \ottnt{b'}  \;  \ottnt{a'}  }{%
{\ottdrulename{E\_App}}{}%
}}

\newcommand{\ottdruleEXXAbsurd}[1]{\ottdrule[#1]{%
\ottpremise{ \ottnt{b}   \equiv   \ottnt{b'} }%
}{
  \mathsf{absurd}( \ottnt{b} )    \equiv    \mathsf{absurd}( \ottnt{b'} )  }{%
{\ottdrulename{E\_Absurd}}{}%
}}

\newcommand{\ottdefnEquiv}[1]{\begin{ottdefnblock}[#1]{$ \ottnt{a}   \equiv   \ottnt{b} $}{\ottcom{Equivalence}}
\ottusedrule{\ottdruleEXXRefl{}}
\ottusedrule{\ottdruleEXXSym{}}
\ottusedrule{\ottdruleEXXTrans{}}
\ottusedrule{\ottdruleEXXBeta{}}
\ottusedrule{\ottdruleEXXArrow{}}
\ottusedrule{\ottdruleEXXPi{}}
\ottusedrule{\ottdruleEXXAbs{}}
\ottusedrule{\ottdruleEXXApp{}}
\ottusedrule{\ottdruleEXXAbsurd{}}
\end{ottdefnblock}}

\newcommand{\ottdefnsJEq}{
\ottdefnEquiv{}}

\newcommand{\ottdruleGXXEmpty}[1]{\ottdrule[#1]{%
}{
\vdash  \varnothing}{%
{\ottdrulename{G\_Empty}}{}%
}}

\newcommand{\ottdruleGXXCons}[1]{\ottdrule[#1]{%
\ottpremise{\vdash  \Gamma}%
\ottpremise{ \Gamma  \vdash  \ottnt{A}  :^{ \ottnt{k} }   \star  }%
\ottpremise{\mathit{x} \, \not\in \, \mathsf{dom} \, \Gamma}%
}{
\vdash   \Gamma ,   \mathit{x}    :^{ \ottnt{k} }  \ottnt{A}   }{%
{\ottdrulename{G\_Cons}}{}%
}}

\newcommand{\ottdefnCtx}[1]{\begin{ottdefnblock}[#1]{$\vdash  \Gamma$}{\ottcom{Context formation}}
\ottusedrule{\ottdruleGXXEmpty{}}
\ottusedrule{\ottdruleGXXCons{}}
\end{ottdefnblock}}

\newcommand{\ottdruleTXXType}[1]{\ottdrule[#1]{%
\ottpremise{\vdash  \Gamma}%
}{
 \Gamma  \vdash   \star   :^{ \ottnt{k} }   \star  }{%
{\ottdrulename{T\_Type}}{}%
}}

\newcommand{\ottdruleTXXVar}[1]{\ottdrule[#1]{%
\ottpremise{ \mathit{x} \! :\!^ \ottnt{j} \ottnt{A}  \in  \Gamma }%
\ottpremise{\vdash  \Gamma}%
\ottpremise{\ottnt{j}  \leq  \ottnt{k}}%
}{
 \Gamma  \vdash   \mathit{ \mathit{x} }   :^{ \ottnt{k} }  \ottnt{A} }{%
{\ottdrulename{T\_Var}}{}%
}}

\newcommand{\ottdruleTXXArrow}[1]{\ottdrule[#1]{%
\ottpremise{ \Gamma  \vdash  \ottnt{A}  :^{ \ottnt{k} }   \star  }%
\ottpremise{ \Gamma  \vdash  \ottnt{B}  :^{ \ottnt{k} }   \star  }%
}{
 \Gamma  \vdash   \ottnt{A}  \mathrel{\rightarrow}  \ottnt{B}   :^{ \ottnt{k} }   \star  }{%
{\ottdrulename{T\_Arrow}}{}%
}}

\newcommand{\ottdruleTXXPi}[1]{\ottdrule[#1]{%
\ottpremise{ \Gamma  \vdash  \ottnt{A}  :^{ \ottnt{j} }   \star  }%
\ottpremise{  \Gamma ,   \mathit{x}    :^{ \ottnt{j} }  \ottnt{A}     \vdash  \ottnt{B}  :^{ \ottnt{k} }   \star  }%
\ottpremise{\ottnt{j}  \ottsym{<}  \ottnt{k}}%
}{
 \Gamma  \vdash   \Pi \mathit{ \mathit{x} }\!:^{ \ottnt{j} }\! \ottnt{A}  \mathpunct{.}  \ottnt{B}   :^{ \ottnt{k} }   \star  }{%
{\ottdrulename{T\_Pi}}{}%
}}

\newcommand{\ottdruleTXXAbsTm}[1]{\ottdrule[#1]{%
\ottpremise{ \Gamma  \vdash  \ottnt{A}  :^{ \ottnt{k} }   \star  }%
\ottpremise{ \Gamma  \vdash  \ottnt{B}  :^{ \ottnt{k} }   \star  }%
\ottpremise{  \Gamma ,   \mathit{x}    :^{ \ottnt{k} }  \ottnt{A}     \vdash  \ottnt{b}  :^{ \ottnt{k} }  \ottnt{B} }%
}{
 \Gamma  \vdash   \lambda \mathit{ \mathit{x} } \mathpunct{.}  \ottnt{b}   :^{ \ottnt{k} }   \ottnt{A}  \mathrel{\rightarrow}  \ottnt{B}  }{%
{\ottdrulename{T\_AbsTm}}{}%
}}

\newcommand{\ottdruleTXXAbsTy}[1]{\ottdrule[#1]{%
\ottpremise{ \Gamma  \vdash  \ottnt{A}  :^{ \ottnt{j} }   \star  }%
\ottpremise{  \Gamma ,   \mathit{x}    :^{ \ottnt{j} }  \ottnt{A}     \vdash  \ottnt{b}  :^{ \ottnt{k} }  \ottnt{B} }%
\ottpremise{\ottnt{j}  \ottsym{<}  \ottnt{k}}%
}{
 \Gamma  \vdash   \lambda \mathit{ \mathit{x} } \mathpunct{.}  \ottnt{b}   :^{ \ottnt{k} }   \Pi \mathit{ \mathit{x} }\!:^{ \ottnt{j} }\! \ottnt{A}  \mathpunct{.}  \ottnt{B}  }{%
{\ottdrulename{T\_AbsTy}}{}%
}}

\newcommand{\ottdruleTXXAppTm}[1]{\ottdrule[#1]{%
\ottpremise{ \Gamma  \vdash  \ottnt{b}  :^{ \ottnt{k} }   \ottnt{A}  \mathrel{\rightarrow}  \ottnt{B}  }%
\ottpremise{ \Gamma  \vdash  \ottnt{a}  :^{ \ottnt{k} }  \ottnt{A} }%
}{
 \Gamma  \vdash   \ottnt{b}  \;  \ottnt{a}   :^{ \ottnt{k} }  \ottnt{B} }{%
{\ottdrulename{T\_AppTm}}{}%
}}

\newcommand{\ottdruleTXXAppTy}[1]{\ottdrule[#1]{%
\ottpremise{ \Gamma  \vdash  \ottnt{b}  :^{ \ottnt{k} }   \Pi \mathit{ \mathit{x} }\!:^{ \ottnt{j} }\! \ottnt{A}  \mathpunct{.}  \ottnt{B}  }%
\ottpremise{ \Gamma  \vdash  \ottnt{a}  :^{ \ottnt{j} }  \ottnt{A} }%
}{
 \Gamma  \vdash   \ottnt{b}  \;  \ottnt{a}   :^{ \ottnt{k} }   \ottnt{B}  \mathopen{ \{ }  \ottnt{a}  / \mathit{ \mathit{x} } \mathclose{ \} }  }{%
{\ottdrulename{T\_AppTy}}{}%
}}

\newcommand{\ottdruleTXXBottom}[1]{\ottdrule[#1]{%
\ottpremise{\vdash  \Gamma}%
}{
 \Gamma  \vdash   \bot   :^{ \ottnt{k} }   \star  }{%
{\ottdrulename{T\_Bottom}}{}%
}}

\newcommand{\ottdruleTXXAbsurd}[1]{\ottdrule[#1]{%
\ottpremise{ \Gamma  \vdash  \ottnt{A}  :^{ \ottnt{k} }   \star  }%
\ottpremise{ \Gamma  \vdash  \ottnt{b}  :^{ \ottnt{k} }   \bot  }%
}{
 \Gamma  \vdash   \mathsf{absurd}( \ottnt{b} )   :^{ \ottnt{k} }  \ottnt{A} }{%
{\ottdrulename{T\_Absurd}}{}%
}}

\newcommand{\ottdruleTXXCumul}[1]{\ottdrule[#1]{%
\ottpremise{ \Gamma  \vdash  \ottnt{a}  :^{ \ottnt{k} }  \ottnt{A} }%
\ottpremise{ \Gamma  \vdash  \ottnt{B}  :^{ \ottnt{k} }   \star  }%
\ottpremise{ \ottnt{A}   \equiv   \ottnt{B} }%
\ottpremise{\vdash  \Gamma}%
}{
 \Gamma  \vdash  \ottnt{a}  :^{ \ottnt{k} }  \ottnt{B} }{%
{\ottdrulename{T\_Cumul}}{}%
}}

\newcommand{\ottdefnTyping}[1]{\begin{ottdefnblock}[#1]{$ \Gamma  \vdash  \ottnt{a}  :^{ \ottnt{k} }  \ottnt{A} $}{\ottcom{Term formation}}
\ottusedrule{\ottdruleTXXType{}}
\ottusedrule{\ottdruleTXXVar{}}
\ottusedrule{\ottdruleTXXArrow{}}
\ottusedrule{\ottdruleTXXPi{}}
\ottusedrule{\ottdruleTXXAbsTm{}}
\ottusedrule{\ottdruleTXXAbsTy{}}
\ottusedrule{\ottdruleTXXAppTm{}}
\ottusedrule{\ottdruleTXXAppTy{}}
\ottusedrule{\ottdruleTXXBottom{}}
\ottusedrule{\ottdruleTXXAbsurd{}}
\ottusedrule{\ottdruleTXXCumul{}}
\end{ottdefnblock}}

\newcommand{\ottdefnsJTyping}{
\ottdefnCtx{}\ottdefnTyping{}}

\newcommand{\ottdruleUXXAbsTy}[1]{\ottdrule[#1]{%
\ottpremise{ \Gamma  \vdash  \ottnt{A}  :^{ \ottnt{j} }   \star  }%
\ottpremise{  \Gamma ,   \mathit{x}    :^{ \ottnt{j} }  \ottnt{A}     \vdash  \ottnt{b}  :^{ \ottnt{k} }  \ottnt{B} }%
\ottpremise{\ottnt{j}  \ottsym{<}  \ottnt{k}}%
}{
 \Gamma  \vdash   \lambda \mathit{ \mathit{x} } \mathpunct{.}  \ottnt{b}   :^{ \ottnt{k} }   \Pi \mathit{ \mathit{x} }\!:\! \ottnt{A}  \mathpunct{.}  \ottnt{B}  }{%
{\ottdrulename{U\_AbsTy}}{}%
}}

\newcommand{\ottdruleUXXPi}[1]{\ottdrule[#1]{%
\ottpremise{ \Gamma  \vdash  \ottnt{A}  :^{ \ottnt{j} }   \star  }%
\ottpremise{  \Gamma ,   \mathit{x}    :^{ \ottnt{j} }  \ottnt{A}     \vdash  \ottnt{B}  :^{ \ottnt{k} }   \star  }%
\ottpremise{\ottnt{j}  \ottsym{<}  \ottnt{k}}%
}{
 \Gamma  \vdash   \Pi \mathit{ \mathit{x} }\!:\! \ottnt{A}  \mathpunct{.}  \ottnt{B}   :^{ \ottnt{k} }   \star  }{%
{\ottdrulename{U\_Pi}}{}%
}}

\newcommand{\ottdefnFakeTyping}[1]{\begin{ottdefnblock}[#1]{$ \Gamma  \vdash  \ottnt{a}  :^{ \ottnt{k} }  \ottnt{A} $}{\ottcom{Term formation}}
\ottusedrule{\ottdruleUXXAbsTy{}}
\ottusedrule{\ottdruleUXXPi{}}
\end{ottdefnblock}}

\newcommand{\ottdefnsJFakeTyping}{
\ottdefnFakeTyping{}}

\newcommand{\ottdruleSXXNil}[1]{\ottdrule[#1]{%
}{
 \varnothing   \leq   \varnothing }{%
{\ottdrulename{S\_Nil}}{}%
}}

\newcommand{\ottdruleSXXCons}[1]{\ottdrule[#1]{%
\ottpremise{\ottnt{j}  \leq  \ottnt{k}}%
\ottpremise{ \Gamma_{{\mathrm{1}}}   \leq   \Gamma_{{\mathrm{2}}} }%
}{
  \Gamma_{{\mathrm{1}}} ,   \mathit{x}    :^{ \ottnt{j} }  \ottnt{A}      \leq    \Gamma_{{\mathrm{2}}} ,   \mathit{x}    :^{ \ottnt{k} }  \ottnt{A}    }{%
{\ottdrulename{S\_Cons}}{}%
}}

\newcommand{\ottdruleSXXWeak}[1]{\ottdrule[#1]{%
\ottpremise{ \Gamma_{{\mathrm{1}}}   \leq   \Gamma_{{\mathrm{2}}} }%
}{
  \Gamma_{{\mathrm{1}}} ,   \mathit{x}    :^{ \ottnt{k} }  \ottnt{A}      \leq   \Gamma_{{\mathrm{2}}} }{%
{\ottdrulename{S\_Weak}}{}%
}}

\newcommand{\ottdefnCtxSub}[1]{\begin{ottdefnblock}[#1]{$ \Gamma_{{\mathrm{1}}}   \leq   \Gamma_{{\mathrm{2}}} $}{}
\ottusedrule{\ottdruleSXXNil{}}
\ottusedrule{\ottdruleSXXCons{}}
\ottusedrule{\ottdruleSXXWeak{}}
\end{ottdefnblock}}

\newcommand{\ottdefnsJSub}{
\ottdefnCtxSub{}}

\newcommand{\ottdruleRXXBeta}[1]{\ottdrule[#1]{%
}{
 \Delta  \vdash   \ottsym{(}   \lambda \mathit{ \mathit{x} } \mathpunct{.}  \ottnt{b}   \ottsym{)}  \;  \ottnt{a}   \leadsto   \ottnt{b}  \mathopen{ \{ }  \ottnt{a}  / \mathit{ \mathit{x} } \mathclose{ \} }  }{%
{\ottdrulename{R\_Beta}}{}%
}}

\newcommand{\ottdruleRXXDelta}[1]{\ottdrule[#1]{%
\ottpremise{ \mathit{x} \! :\!^ \ottnt{k} \ottnt{A}  \coloneqq  \ottnt{a}  \in  \Delta }%
}{
 \Delta  \vdash   \mathit{ \mathit{x} }^{ \ottnt{i} }   \leadsto   \ottnt{a} ^{+ \ottnt{i} }  }{%
{\ottdrulename{R\_Delta}}{}%
}}

\newcommand{\ottdruleRXXApp}[1]{\ottdrule[#1]{%
\ottpremise{ \Delta  \vdash  \ottnt{b}  \leadsto  \ottnt{b'} }%
}{
 \Delta  \vdash   \ottnt{b}  \;  \ottnt{a}   \leadsto   \ottnt{b'}  \;  \ottnt{a}  }{%
{\ottdrulename{R\_App}}{}%
}}

\newcommand{\ottdruleRXXAbsurd}[1]{\ottdrule[#1]{%
\ottpremise{ \Delta  \vdash  \ottnt{b}  \leadsto  \ottnt{b'} }%
}{
 \Delta  \vdash   \mathsf{absurd}( \ottnt{b} )   \leadsto   \mathsf{absurd}( \ottnt{b'} )  }{%
{\ottdrulename{R\_Absurd}}{}%
}}

\newcommand{\ottdefnReduce}[1]{\begin{ottdefnblock}[#1]{$ \Delta  \vdash  \ottnt{a}  \leadsto  \ottnt{b} $}{\ottcom{Single-step reduction}}
\ottusedrule{\ottdruleRXXBeta{}}
\ottusedrule{\ottdruleRXXDelta{}}
\ottusedrule{\ottdruleRXXApp{}}
\ottusedrule{\ottdruleRXXAbsurd{}}
\end{ottdefnblock}}

\newcommand{\ottdefnsAReduce}{
\ottdefnReduce{}}

\newcommand{\ottdruleWXXRefl}[1]{\ottdrule[#1]{%
}{
 \Delta  \vdash  \ottnt{a}  \leadsto^{\ast}  \ottnt{a} }{%
{\ottdrulename{W\_Refl}}{}%
}}

\newcommand{\ottdruleWXXTrans}[1]{\ottdrule[#1]{%
\ottpremise{ \Delta  \vdash  \ottnt{a}  \leadsto  \ottnt{b} }%
\ottpremise{ \Delta  \vdash  \ottnt{b}  \leadsto^{\ast}  \ottnt{c} }%
}{
 \Delta  \vdash  \ottnt{a}  \leadsto^{\ast}  \ottnt{c} }{%
{\ottdrulename{W\_Trans}}{}%
}}

\newcommand{\ottdefnWHNF}[1]{\begin{ottdefnblock}[#1]{$ \Delta  \vdash  \ottnt{a}  \leadsto^{\ast}  \ottnt{b} $}{\ottcom{Weak head normalization}}
\ottusedrule{\ottdruleWXXRefl{}}
\ottusedrule{\ottdruleWXXTrans{}}
\end{ottdefnblock}}

\newcommand{\ottdefnsAWHNF}{
\ottdefnWHNF{}}

\newcommand{\ottdruleDEXXRefl}[1]{\ottdrule[#1]{%
}{
 \Delta  \vdash  \ottnt{a} \equiv \ottnt{a} }{%
{\ottdrulename{DE\_Refl}}{}%
}}

\newcommand{\ottdruleDEXXSym}[1]{\ottdrule[#1]{%
\ottpremise{ \Delta  \vdash  \ottnt{b} \equiv \ottnt{a} }%
}{
 \Delta  \vdash  \ottnt{a} \equiv \ottnt{b} }{%
{\ottdrulename{DE\_Sym}}{}%
}}

\newcommand{\ottdruleDEXXTrans}[1]{\ottdrule[#1]{%
\ottpremise{ \Delta  \vdash  \ottnt{a} \equiv \ottnt{b} }%
\ottpremise{ \Delta  \vdash  \ottnt{b} \equiv \ottnt{c} }%
}{
 \Delta  \vdash  \ottnt{a} \equiv \ottnt{c} }{%
{\ottdrulename{DE\_Trans}}{}%
}}

\newcommand{\ottdruleDEXXBeta}[1]{\ottdrule[#1]{%
}{
 \Delta  \vdash   \ottsym{(}   \lambda \mathit{ \mathit{x} } \mathpunct{.}  \ottnt{b}   \ottsym{)}  \;  \ottnt{a}  \equiv  \ottnt{b}  \mathopen{ \{ }  \ottnt{a}  / \mathit{ \mathit{x} } \mathclose{ \} }  }{%
{\ottdrulename{DE\_Beta}}{}%
}}

\newcommand{\ottdruleDEXXDelta}[1]{\ottdrule[#1]{%
\ottpremise{ \mathit{x} \! :\!^ \ottnt{k} \ottnt{A}  \coloneqq  \ottnt{a}  \in  \Delta }%
\ottpremise{  }%
}{
 \Delta  \vdash   \mathit{ \mathit{x} }^{ \ottnt{i} }  \equiv  \ottnt{a} ^{+ \ottnt{i} }  }{%
{\ottdrulename{DE\_Delta}}{}%
}}

\newcommand{\ottdruleDEXXArrow}[1]{\ottdrule[#1]{%
\ottpremise{ \Delta  \vdash  \ottnt{A} \equiv \ottnt{A'} }%
\ottpremise{ \Delta  \vdash  \ottnt{B} \equiv \ottnt{B'} }%
}{
 \Delta  \vdash   \ottnt{A}  \mathrel{\rightarrow}  \ottnt{B}  \equiv  \ottnt{A'}  \mathrel{\rightarrow}  \ottnt{B'}  }{%
{\ottdrulename{DE\_Arrow}}{}%
}}

\newcommand{\ottdruleDEXXPi}[1]{\ottdrule[#1]{%
\ottpremise{ \Delta  \vdash  \ottnt{A} \equiv \ottnt{A'} }%
\ottpremise{ \Delta  \vdash  \ottnt{B} \equiv \ottnt{B'} }%
}{
 \Delta  \vdash   \Pi \mathit{ \mathit{x} }\!:^{ \ottnt{k} }\! \ottnt{A}  \mathpunct{.}  \ottnt{B}  \equiv  \Pi \mathit{ \mathit{x} }\!:^{ \ottnt{k} }\! \ottnt{A'}  \mathpunct{.}  \ottnt{B'}  }{%
{\ottdrulename{DE\_Pi}}{}%
}}

\newcommand{\ottdruleDEXXAbs}[1]{\ottdrule[#1]{%
\ottpremise{ \Delta  \vdash  \ottnt{b} \equiv \ottnt{b'} }%
}{
 \Delta  \vdash   \lambda \mathit{ \mathit{x} } \mathpunct{.}  \ottnt{b}  \equiv  \lambda \mathit{ \mathit{x} } \mathpunct{.}  \ottnt{b'}  }{%
{\ottdrulename{DE\_Abs}}{}%
}}

\newcommand{\ottdruleDEXXApp}[1]{\ottdrule[#1]{%
\ottpremise{ \Delta  \vdash  \ottnt{a} \equiv \ottnt{a'} }%
\ottpremise{ \Delta  \vdash  \ottnt{b} \equiv \ottnt{b'} }%
}{
 \Delta  \vdash   \ottnt{b}  \;  \ottnt{a}  \equiv  \ottnt{b'}  \;  \ottnt{a'}  }{%
{\ottdrulename{DE\_App}}{}%
}}

\newcommand{\ottdruleDEXXAbsurd}[1]{\ottdrule[#1]{%
\ottpremise{ \Delta  \vdash  \ottnt{b} \equiv \ottnt{b'} }%
}{
 \Delta  \vdash   \mathsf{absurd}( \ottnt{b} )  \equiv  \mathsf{absurd}( \ottnt{b'} )  }{%
{\ottdrulename{DE\_Absurd}}{}%
}}

\newcommand{\ottdruleDEXXEtaAlt}[1]{\ottdrule[#1]{%
}{
 \Delta  \vdash   \lambda \mathit{ \mathit{x} } \mathpunct{.}    \ottnt{b}  \;   \mathit{ \mathit{x} }     \equiv \ottnt{b} }{%
{\ottdrulename{DE\_EtaAlt}}{}%
}}

\newcommand{\ottdefnDEquiv}[1]{\begin{ottdefnblock}[#1]{$ \Delta  \vdash  \ottnt{a} \equiv \ottnt{b} $}{\ottcom{Equivalence}}
\ottusedrule{\ottdruleDEXXRefl{}}
\ottusedrule{\ottdruleDEXXSym{}}
\ottusedrule{\ottdruleDEXXTrans{}}
\ottusedrule{\ottdruleDEXXBeta{}}
\ottusedrule{\ottdruleDEXXDelta{}}
\ottusedrule{\ottdruleDEXXArrow{}}
\ottusedrule{\ottdruleDEXXPi{}}
\ottusedrule{\ottdruleDEXXAbs{}}
\ottusedrule{\ottdruleDEXXApp{}}
\ottusedrule{\ottdruleDEXXAbsurd{}}
\ottusedrule{\ottdruleDEXXEtaAlt{}}
\end{ottdefnblock}}

\newcommand{\ottdefnsJDEq}{
\ottdefnDEquiv{}}

\newcommand{\ottdruleDXXEmpty}[1]{\ottdrule[#1]{%
}{
\vdash  \varnothing}{%
{\ottdrulename{D\_Empty}}{}%
}}

\newcommand{\ottdruleDXXCons}[1]{\ottdrule[#1]{%
\ottpremise{\vdash  \Delta}%
\ottpremise{ \Delta ;  \varnothing  \vdash  \ottnt{A}  :^{ \ottnt{k} }   \star  }%
\ottpremise{ \Delta ;  \varnothing  \vdash  \ottnt{a}  :^{ \ottnt{k} }  \ottnt{A} }%
\ottpremise{\mathit{x} \, \not\in \, \mathsf{dom} \, \Delta}%
}{
\vdash   \Delta ,   \mathit{x}    :^{ \ottnt{k} }  \ottnt{A}  \coloneqq  \ottnt{a}   }{%
{\ottdrulename{D\_Cons}}{}%
}}

\newcommand{\ottdefnDSig}[1]{\begin{ottdefnblock}[#1]{$\vdash  \Delta$}{\ottcom{Signature well-formedness}}
\ottusedrule{\ottdruleDXXEmpty{}}
\ottusedrule{\ottdruleDXXCons{}}
\end{ottdefnblock}}

\newcommand{\ottdruleDGXXEmpty}[1]{\ottdrule[#1]{%
\ottpremise{\vdash  \Delta}%
}{
\Delta  \vdash  \varnothing}{%
{\ottdrulename{DG\_Empty}}{}%
}}

\newcommand{\ottdruleDGXXCons}[1]{\ottdrule[#1]{%
\ottpremise{\Delta  \vdash  \Gamma}%
\ottpremise{ \Delta ;  \Gamma  \vdash  \ottnt{A}  :^{ \ottnt{k} }   \star  }%
\ottpremise{\mathit{x} \, \not\in \, \mathsf{dom} \, \Gamma}%
\ottpremise{\mathit{x} \, \not\in \, \mathsf{dom} \, \Delta}%
}{
\Delta  \vdash   \Gamma ,   \mathit{x}    :^{ \ottnt{k} }  \ottnt{A}   }{%
{\ottdrulename{DG\_Cons}}{}%
}}

\newcommand{\ottdefnDCtx}[1]{\begin{ottdefnblock}[#1]{$\Delta  \vdash  \Gamma$}{\ottcom{Context formation}}
\ottusedrule{\ottdruleDGXXEmpty{}}
\ottusedrule{\ottdruleDGXXCons{}}
\end{ottdefnblock}}

\newcommand{\ottdruleDTXXType}[1]{\ottdrule[#1]{%
\ottpremise{\Delta  \vdash  \Gamma}%
}{
 \Delta ;  \Gamma  \vdash   \star   :^{ \ottnt{k} }   \star  }{%
{\ottdrulename{DT\_Type}}{}%
}}

\newcommand{\ottdruleDTXXConst}[1]{\ottdrule[#1]{%
\ottpremise{ \mathit{x} \! :\!^ \ottnt{j} \ottnt{A}  \coloneqq  \ottnt{a}  \in  \Delta }%
\ottpremise{\Delta  \vdash  \Gamma}%
\ottpremise{\vdash  \Delta}%
\ottpremise{\ottnt{i}  \ottsym{+}  \ottnt{j}  \leq  \ottnt{k}}%
}{
 \Delta ;  \Gamma  \vdash   \mathit{ \mathit{x} }^{ \ottnt{i} }   :^{ \ottnt{k} }   \ottnt{A} ^{+ \ottnt{i} }  }{%
{\ottdrulename{DT\_Const}}{}%
}}

\newcommand{\ottdruleDTXXVar}[1]{\ottdrule[#1]{%
\ottpremise{ \mathit{x} \! :\!^ \ottnt{j} \ottnt{A}  \in  \Gamma }%
\ottpremise{\Delta  \vdash  \Gamma}%
\ottpremise{\ottnt{j}  \leq  \ottnt{k}}%
}{
 \Delta ;  \Gamma  \vdash   \mathit{ \mathit{x} }   :^{ \ottnt{k} }  \ottnt{A} }{%
{\ottdrulename{DT\_Var}}{}%
}}

\newcommand{\ottdruleDTXXArrow}[1]{\ottdrule[#1]{%
\ottpremise{ \Delta ;  \Gamma  \vdash  \ottnt{A}  :^{ \ottnt{k} }   \star  }%
\ottpremise{ \Delta ;  \Gamma  \vdash  \ottnt{B}  :^{ \ottnt{k} }   \star  }%
}{
 \Delta ;  \Gamma  \vdash   \ottnt{A}  \mathrel{\rightarrow}  \ottnt{B}   :^{ \ottnt{k} }   \star  }{%
{\ottdrulename{DT\_Arrow}}{}%
}}

\newcommand{\ottdruleDTXXPi}[1]{\ottdrule[#1]{%
\ottpremise{ \Delta ;  \Gamma  \vdash  \ottnt{A}  :^{ \ottnt{j} }   \star  }%
\ottpremise{ \Delta ;   \Gamma ,   \mathit{x}    :^{ \ottnt{j} }  \ottnt{A}     \vdash  \ottnt{B}  :^{ \ottnt{k} }   \star  }%
\ottpremise{\ottnt{j}  \ottsym{<}  \ottnt{k}}%
}{
 \Delta ;  \Gamma  \vdash   \Pi \mathit{ \mathit{x} }\!:^{ \ottnt{j} }\! \ottnt{A}  \mathpunct{.}  \ottnt{B}   :^{ \ottnt{k} }   \star  }{%
{\ottdrulename{DT\_Pi}}{}%
}}

\newcommand{\ottdruleDTXXAbsTm}[1]{\ottdrule[#1]{%
\ottpremise{ \Delta ;  \Gamma  \vdash  \ottnt{A}  :^{ \ottnt{k} }   \star  }%
\ottpremise{ \Delta ;  \Gamma  \vdash  \ottnt{B}  :^{ \ottnt{k} }   \star  }%
\ottpremise{ \Delta ;   \Gamma ,   \mathit{x}    :^{ \ottnt{k} }  \ottnt{A}     \vdash  \ottnt{b}  :^{ \ottnt{k} }  \ottnt{B} }%
}{
 \Delta ;  \Gamma  \vdash   \lambda \mathit{ \mathit{x} } \mathpunct{.}  \ottnt{b}   :^{ \ottnt{k} }   \ottnt{A}  \mathrel{\rightarrow}  \ottnt{B}  }{%
{\ottdrulename{DT\_AbsTm}}{}%
}}

\newcommand{\ottdruleDTXXAbsTy}[1]{\ottdrule[#1]{%
\ottpremise{ \Delta ;  \Gamma  \vdash  \ottnt{A}  :^{ \ottnt{j} }   \star  }%
\ottpremise{ \Delta ;   \Gamma ,   \mathit{x}    :^{ \ottnt{j} }  \ottnt{A}     \vdash  \ottnt{b}  :^{ \ottnt{k} }  \ottnt{B} }%
\ottpremise{\ottnt{j}  \ottsym{<}  \ottnt{k}}%
}{
 \Delta ;  \Gamma  \vdash   \lambda \mathit{ \mathit{x} } \mathpunct{.}  \ottnt{b}   :^{ \ottnt{k} }   \Pi \mathit{ \mathit{x} }\!:^{ \ottnt{j} }\! \ottnt{A}  \mathpunct{.}  \ottnt{B}  }{%
{\ottdrulename{DT\_AbsTy}}{}%
}}

\newcommand{\ottdruleDTXXAppTm}[1]{\ottdrule[#1]{%
\ottpremise{ \Delta ;  \Gamma  \vdash  \ottnt{b}  :^{ \ottnt{k} }   \ottnt{A}  \mathrel{\rightarrow}  \ottnt{B}  }%
\ottpremise{ \Delta ;  \Gamma  \vdash  \ottnt{a}  :^{ \ottnt{k} }  \ottnt{A} }%
}{
 \Delta ;  \Gamma  \vdash   \ottnt{b}  \;  \ottnt{a}   :^{ \ottnt{k} }  \ottnt{B} }{%
{\ottdrulename{DT\_AppTm}}{}%
}}

\newcommand{\ottdruleDTXXAppTy}[1]{\ottdrule[#1]{%
\ottpremise{ \Delta ;  \Gamma  \vdash  \ottnt{b}  :^{ \ottnt{k} }   \Pi \mathit{ \mathit{x} }\!:^{ \ottnt{j} }\! \ottnt{A}  \mathpunct{.}  \ottnt{B}  }%
\ottpremise{ \Delta ;  \Gamma  \vdash  \ottnt{a}  :^{ \ottnt{j} }  \ottnt{A} }%
\ottpremise{\ottnt{j}  \ottsym{<}  \ottnt{k}}%
}{
 \Delta ;  \Gamma  \vdash   \ottnt{b}  \;  \ottnt{a}   :^{ \ottnt{k} }   \ottnt{B}  \mathopen{ \{ }  \ottnt{a}  / \mathit{ \mathit{x} } \mathclose{ \} }  }{%
{\ottdrulename{DT\_AppTy}}{}%
}}

\newcommand{\ottdruleDTXXBottom}[1]{\ottdrule[#1]{%
\ottpremise{\Delta  \vdash  \Gamma}%
}{
 \Delta ;  \Gamma  \vdash   \bot   :^{ \ottnt{k} }   \star  }{%
{\ottdrulename{DT\_Bottom}}{}%
}}

\newcommand{\ottdruleDTXXAbsurd}[1]{\ottdrule[#1]{%
\ottpremise{ \Delta ;  \Gamma  \vdash  \ottnt{A}  :^{ \ottnt{k} }   \star  }%
\ottpremise{ \Delta ;  \Gamma  \vdash  \ottnt{b}  :^{ \ottnt{k} }   \bot  }%
}{
 \Delta ;  \Gamma  \vdash   \mathsf{absurd}( \ottnt{b} )   :^{ \ottnt{k} }  \ottnt{A} }{%
{\ottdrulename{DT\_Absurd}}{}%
}}

\newcommand{\ottdruleDTXXConv}[1]{\ottdrule[#1]{%
\ottpremise{ \Delta ;  \Gamma  \vdash  \ottnt{a}  :^{ \ottnt{k} }  \ottnt{A} }%
\ottpremise{ \Delta ;  \Gamma  \vdash  \ottnt{B}  :^{ \ottnt{k} }   \star  }%
\ottpremise{ \Delta  \vdash  \ottnt{A} \equiv \ottnt{B} }%
}{
 \Delta ;  \Gamma  \vdash  \ottnt{a}  :^{ \ottnt{k} }  \ottnt{B} }{%
{\ottdrulename{DT\_Conv}}{}%
}}

\newcommand{\ottdruleDTXXCumul}[1]{\ottdrule[#1]{%
\ottpremise{ \Delta ;  \Gamma  \vdash  \ottnt{a}  :^{ \ottnt{j} }  \ottnt{A} }%
\ottpremise{\ottnt{j}  \leq  \ottnt{k}}%
}{
 \Delta ;  \Gamma  \vdash  \ottnt{a}  :^{ \ottnt{k} }  \ottnt{A} }{%
{\ottdrulename{DT\_Cumul}}{}%
}}

\newcommand{\ottdefnDTyping}[1]{\begin{ottdefnblock}[#1]{$ \Delta ;  \Gamma  \vdash  \ottnt{a}  :^{ \ottnt{k} }  \ottnt{A} $}{\ottcom{Term formation}}
\ottusedrule{\ottdruleDTXXType{}}
\ottusedrule{\ottdruleDTXXConst{}}
\ottusedrule{\ottdruleDTXXVar{}}
\ottusedrule{\ottdruleDTXXArrow{}}
\ottusedrule{\ottdruleDTXXPi{}}
\ottusedrule{\ottdruleDTXXAbsTm{}}
\ottusedrule{\ottdruleDTXXAbsTy{}}
\ottusedrule{\ottdruleDTXXAppTm{}}
\ottusedrule{\ottdruleDTXXAppTy{}}
\ottusedrule{\ottdruleDTXXBottom{}}
\ottusedrule{\ottdruleDTXXAbsurd{}}
\ottusedrule{\ottdruleDTXXConv{}}
\ottusedrule{\ottdruleDTXXCumul{}}
\end{ottdefnblock}}

\newcommand{\ottdefnsJDTyping}{
\ottdefnDSig{}\ottdefnDCtx{}\ottdefnDTyping{}}

\newcommand{\ottdefnss}{
\ottdefnsJEq
\ottdefnsJTyping
\ottdefnsJFakeTyping
\ottdefnsJSub
\ottdefnsAReduce
\ottdefnsAWHNF
\ottdefnsJDEq
\ottdefnsJDTyping
}

\newcommand{\ottall}{\ottmetavars\\[0pt]
\ottgrammar\\[5.0mm]
\ottdefnss}

  \renewottcommands[ott]

\newcommand\suppress[1]{}

\section{Introduction}

Ever since their introduction in Martin-L\"of's intuitionistic type theory (MLTT)~\citep{mltt},
dependent type theories have included hierarchies of type universes in order to
rectify the logical inconsistency of the type-in-type axiom.
That is, rather than the universe $ \star $ of types being its own type,
these type theories have universes $ \star _{k}$ indexed by a sequence of levels $k$
such that the type of a universe is the universe at the next higher level.

Such a universe hierarchy is a rudimentary ingredient in many contemporary proof assistants,
such as Coq \citep{coq}, Agda \citep{agda}, Lean \citep{lean},
F* \citep{fstar}, and Arend \citep{arend}.
For greater expressiveness, all of these
also implement some sort of level polymorphism.
Supporting such generality means that the proof assistant must handle
level variable constraints, level expressions, or both.
However, programming with and especially debugging errors involving universe levels
is a common pain point among proof assistant users.
\jcxz{[\sc{citation needed}]}
So we ask: do all roads necessarily lead to level polymorphism
and more generally a universe hierarchy,
or are there other avenues to be taken?

In this work, we design \textbf{\langue} (\lang)
to explore one potential alternative:
rather than stratifying universes into a hierarchy,
we instead stratify \emph{typing judgements} themselves by levels.
This is inspired by Leivant's \emph{Stratified System F}~\citep{stratified-systemf},
a predicative variant of System F~\citep{girard:systemf,reynolds:systemf}.
Recall the formation rule for polymorphic type quantification in System F,
given below on the left.
The quantification is said to be \emph{impredicative}
because it quantifies over all types including itself,
and so the type $\forall x \mathpunct{.} B$ itself can be substituted for $x$ in $B$.
\begin{mathpar}
  \inferrule*[Lab=F-impredicative]{\Gamma, x \type{} \vdash B \type}{\Gamma \vdash \forall x \mathpunct{.} B \type}
  \and
  \inferrule*[Lab=F-stratified]{
    \Gamma, x \type j \vdash B \type k \\
    j < k
  }{
    \Gamma \vdash \forall x^j \mathpunct{.} B \type k
  }
\end{mathpar}

In contrast, the formation rule in Stratified System F above on the right
disallows impredicativity by restricting polymorphic quantification
to only types that are well formed at strictly lower stratification levels,
and type \wellformedness judgements are additionally indexed by a level.

To extend stratified polymorphism to dependent types,
there are two ways to read this judgement form.
We could interpret $\Gamma \vdash A \type k$ as a type $A$
living in some stratified type universe $ \star _k$,
which would correspond to a usual predicative type theory
where $ \star _j :  \star _k$ when $j < k$.
Alternatively, we can continue to interpret the level $k$ as a property of the judgement
and generalize it to the judgement form $ \Gamma  \vdash  \ottnt{a}  :^{ \ottnt{k} }  \ottnt{A} $,
where variables $ \mathit{x}    :^{ \ottnt{k} }  \ottnt{A}  $ are also annotated with a level within the context $\Gamma$.
Guided by these principles, we introduce stratified dependent function types $ \Pi \mathit{ \mathit{x} }\!:^{ \ottnt{j} }\! \ottnt{A}  \mathpunct{.}  \ottnt{B} $,
which similarly quantify over types at the annotated level $\ottnt{j}$
that must be strictly lower than the overall level of the type.

To enable code reuse, in place of level polymorphism,
we employ McBride's \emph{\crude}~\citep{crude}.
Following Favonia, Angiuli, and Mullanix~\citep{displacement},
we refer to this as \emph{displacement} to prevent confusion.
Given some signature $\Delta$ of global definitions,
we are permitted to use any definition with all of its levels uniformly displaced upwards.

However, even in the presence of displacement,
we find that stratification is sometimes \emph{too} restrictive
and can rule out terms that are otherwise typeable in an unstratified system.
Therefore, \lang includes a separate unstratified nondependent function type with a \emph{floating} domain,
so called because of its behaviour in the presence of cumulativity with respect to the levels.
For a dependent function type, cumulativity can raise its overall level,
but the level of the domain type remains fixed due to its level annotation.
For a floating, nondependent function type whose level is raised by cumulativity,
the domain type here instead floats to have the same level.

In the absence of floating nondependent functions,
with only stratified dependent functions,
consistency holds even with type-in-type,
because the restriction on the domains of dependent functions
prevents the kind of self-referential trickery that permits the usual paradoxes.
However, we haven't yet proven consistency
with the inclusion of floating nondependent functions;
the primary barrier is the covariant behaviour of the floating domain with respect to levels,
which is unusual for function types.
Even so, we have found it impossible to encode some well-known type-theoretic paradoxes,
leading us to believe that consistency \emph{does} hold,
which would make the system suitable as a foundation for theorem proving.

These features form the basis of \lang,
and our contributions are as follows:

\begin{itemize}[leftmargin=\parindent]
  \item A subsystem \sublang,
    featuring only stratified dependent functions and displacement,
    which is then extended to the full \lang with floating nondependent functions.
    $\hookrightarrow$~\Cref{sec:stratt}
  \item A number of examples
    to demonstrate the expressivity of \lang
    and especially to motivate floating functions.
    $\hookrightarrow$~\Cref{sec:examples}
  \item Two major metatheorems:
    logical consistency for \sublang, which is mechanized in Agda,
    and type safety for \lang, which is mechanized in Coq.
    Consistency for the full \lang remains an open problem.
    $\hookrightarrow$~\Cref{sec:metatheory}
  \item A prototype implementation of a type checker,
    which extends \lang to include datatypes
    to demonstrate the effectiveness of stratification and displacement
    in practical dependently-typed programming.
    $\hookrightarrow$~\Cref{sec:implementation}
\end{itemize}

We discuss potential avenues for proving consistency of the full \lang
and compare the useability of its design to existing proof assistants
in terms of working with universe levels in \cref{sec:discussion},
and conclude in \cref{sec:conclusion}.
Our Agda and Coq mechanizations along with the prototype implementation are available
in the supplementary material.
Where lemmas and theorems are first introduced,
we include a footnote indicating the corresponding source file and lemma name in the development.

\section{\langue} \label{sec:stratt}

In this section, we present \langue in two parts.
First is the subsystem \sublang,
which contains the two core features of stratified dependent function types
and global definitions with level displacement.
We then extend it to the full \lang by adding floating nondependent function types.
As the system is fairly small with few parts,
we delay illustrative examples to \cref{sec:examples},
and begin with the formal description.

\subsection{The subsystem \sublang} \label{sec:core}

The subsystem \sublang is a cumulative, extrinsic type theory with types \`a la Russell,
a single type universe, dependent functions, an empty type, and global definitions.
The most significant difference between \sublang and other type theories with these features
is the annotation of the typing judgement with a level
in place of universes in a hierarchy.
We use the naturals and their usual strict order and addition operation for our levels,
but they should be generalizable to any displacement algebra~\citep{displacement}.
The syntax is given below, with $\mathit{x}, \mathit{y}, \mathit{z}$ for variable and constant names
and $\ottnt{i}, \ottnt{j}, \ottnt{k}$ for levels.
\begin{align*}
  \ottnt{a}, \ottnt{b}, \ottnt{c}, \ottnt{A}, \ottnt{B}, \ottnt{C} &\Coloneqq %
     \star  \mid \mathit{x} \mid  \mathit{ \mathit{x} }^{ \ottnt{i} }  \mid  \Pi \mathit{ \mathit{x} }\!:^{ \ottnt{j} }\! \ottnt{A}  \mathpunct{.}  \ottnt{B}  \mid  \lambda \mathit{ \mathit{x} } \mathpunct{.}  \ottnt{b}  \mid  \ottnt{b}  \;  \ottnt{a}  \mid  \bot  \mid  \mathsf{absurd}( \ottnt{b} ) 
\end{align*}

\begin{figure}[h]
  \drules[DT]{$ \Delta ;  \Gamma  \vdash  \ottnt{a}  :^{ \ottnt{k} }  \ottnt{A} $}{Typing}{Type,Pi,AbsTy,AppTy,Var,Const,Bottom,Absurd,Conv}
  \caption{Typing rules (\sublang)}
  \label{fig:substratt-typing}
\end{figure}

The typing judgement has the form
\fbox{$ \Delta ;  \Gamma  \vdash  \ottnt{a}  :^{ \ottnt{k} }  \ottnt{A} $};
its typing rules are given in \cref{fig:substratt-typing}.
The judgement states that term $\ottnt{a}$ is well typed at level $\ottnt{k}$ with type $\ottnt{A}$
under the context $\Gamma$ and signature $\Delta$.
A context consists of declarations $ \mathit{x}    :^{ \ottnt{k} }  \ottnt{A}  $ of variables $\mathit{x}$ of type $\ottnt{A}$ at level $\ottnt{k}$;
variables represent locations where an entire typing derivation may be substituted into the term,
so they also need level annotations.
A signature consists of global definitions $ \mathit{x}    :^{ \ottnt{k} }  \ottnt{A}  \coloneqq  \ottnt{a}  $
of constants $\mathit{x}$ of type $\ottnt{A}$ definitionally equal to $\ottnt{a}$ at level $\ottnt{k}$;
they represent complete typing derivations that will eventually be substituted into the term.

Because stratified judgements replace stratified universes,
the type of the type universe $ \star $ is itself at any level in \rref{DT-Type}.
Stratification is enforced in dependent function types in \rref{DT-Pi}:
the domain type must be well typed at a strictly smaller level
relative to the codomain type and the overall function type.
Similarly, in \rref{DT-AbsTy}, the body of a dependent function
is well typed when its argument and its type are well typed
at a strictly smaller level,
and by \rref{DT-AppTy}, a dependent function
can only be applied to an argument at the strictly smaller domain level.

  Note that the level annotation on dependent function types is necessary for consistency.
  Informally, suppose we have some unannotated type $ \Pi \mathit{ X }\!:\!  \star   \mathpunct{.}  \ottnt{B} $
  and a function of this type, both at level $1$.
  By cumulativity, we can raise the level of the function to $2$,
  then apply it to its own type $ \Pi \mathit{ X }\!:\!  \star   \mathpunct{.}  \ottnt{B} $.
  In short, impredicativity is reintroduced, and stratification defeated.

\Rref{DT-Bottom,DT-Absurd} are the uninhabited type and its eliminator, respectively.
It should be consistent to eliminate a falsehood into any level, including lower levels,
but when viewed bottom-up, the level of the conclusion represents the level of the entire derivation tree,
or the level of all the pieces used to construct the tree,
so it wouldn't make sense to allow premises at higher levels.

In \rref{DT-Var,DT-Const},
variables and constants at level $\ottnt{j}$ can be used at any larger level $\ottnt{k}$,
which we refer to as subsumption.
This permits the following admissible cumulativity lemma,
allowing entire derivations to be used at larger levels.

\begin{lemma}[Cumulativity]\stack{. }\footnote{\footfilethm{coq/restrict.v}{DTyping\_cumul}}
  If $ \Delta ;  \Gamma  \vdash  \ottnt{a}  :^{ \ottnt{j} }  \ottnt{A} $ and $\ottnt{j}  \leq  \ottnt{k}$ then $ \Delta ;  \Gamma  \vdash  \ottnt{a}  :^{ \ottnt{k} }  \ottnt{A} $.
\end{lemma}

Constants are also annotated with a superscript indicating how much they're displaced by.
If a constant $\mathit{x}$ is defined with a type $\ottnt{A}$,
we're permitted to use $ \mathit{ \mathit{x} }^{ \ottnt{i} } $ as an element of type $\ottnt{A}$
but with all of its levels incremented by $\ottnt{i}$.
The metafunction $ \ottnt{a} ^{+ \ottnt{i} } $ performs this increment in the term $\ottnt{a}$,
defined recursively with $ \ottsym{(}   \Pi \mathit{ \mathit{x} }\!:^{ \ottnt{j} }\! \ottnt{A}  \mathpunct{.}  \ottnt{B}   \ottsym{)} ^{+ \ottnt{i} }  =  \Pi \mathit{ \mathit{x} }\!:^{  \ottnt{i}  \ottsym{+}  \ottnt{j}  }\!   \ottnt{A} ^{+ \ottnt{i} }    \mathpunct{.}    \ottnt{B} ^{+ \ottnt{i} }   $
and $ \ottsym{(}   \mathit{ \mathit{x} }^{ \ottnt{j} }   \ottsym{)} ^{+ \ottnt{i} }  =  \mathit{ \mathit{x} }^{  \ottnt{i}  \ottsym{+}  \ottnt{j}  } $.
Constants come from signatures and variables from contexts,
whose key formation rules for the judgements \fbox{$\vdash  \Delta$} and \fbox{$\Delta  \vdash  \Gamma$} respectively are given below.
\begin{gather*}
  \drule[width=0.5\textwidth]{D-Cons}
  \drule[width=0.5\textwidth]{DG-Cons}
\end{gather*}

In \rref{DT-Conv}, we use an untyped definitional equality
\fbox{$ \Delta  \vdash  \ottnt{a} \equiv \ottnt{b} $} that is reflexive, symmetric, transitive, and
congruent.
The full set of rules are given in \cref{fig:wfeq},
including $\beta$-equivalence for functions (\rref{DE-Beta})
and $\delta$-equivalence of constants $\mathit{x}$ with their definitions (\rref{DE-Delta}).
When a constant is displaced as $ \mathit{ \mathit{x} }^{ \ottnt{i} } $,
we must also increment the level annotations in their definitions by $\ottnt{i}$.

\begin{figure}[h]
\drules[DE]{$ \Delta  \vdash  \ottnt{a} \equiv \ottnt{b} $}{Definitional equality}{Refl,Sym,Trans,Beta,Delta,Pi,Abs,App,Absurd}
\caption{Definitional equality rules (\sublang)}
\label{fig:wfeq}
\end{figure}

Given a \welltyped, locally-closed term $ \Delta ;  \varnothing  \vdash  \ottnt{a}  :^{ \ottnt{k} }  \ottnt{A} $,
the entire derivation itself can be displaced upwards by some increment $\ottnt{i}$.
This lemma differs from cumulativity,
since the level annotations in the term and its type are displaced as well,
not just that of the judgement.

\begin{lemma}[Displaceability (empty context)]\stack{. }\footnote{\footfilethm{coq/incr.v}{DTyping\_incr}} \label{lem:displaceability}
  If $ \Delta ;  \varnothing  \vdash  \ottnt{a}  :^{ \ottnt{k} }  \ottnt{A} $ then $ \Delta ;  \varnothing  \vdash   \ottnt{a} ^{+ \ottnt{i} }   :^{  \ottnt{i}  \ottsym{+}  \ottnt{k}  }   \ottnt{A} ^{+ \ottnt{i} }  $.
\end{lemma}

With $ \mathit{x}    :^{ \ottnt{k} }  \ottnt{A}  \coloneqq  \ottnt{a}  $ in the signature,
$ \mathit{ \mathit{x} }^{ \ottnt{i} } $ is definitionally equal to $ \ottnt{a} ^{+ \ottnt{i} } $,
so this lemma justifies \rref{DT-Const},
which would give this displaced constant the type $ \ottnt{A} ^{+ \ottnt{i} } $.

\subsection{Floating functions} \label{sec:floating}

As we'll see in the next section, \sublang alone is insufficiently expressive,
with some examples being unexpectedly untypeable
and others being simply clunky to work with
as a result of the strict restriction on function domains.
The full \lang system therefore extends the subsystem with a separate nondependent function type, 
written $ \ottnt{A}  \mathrel{\rightarrow}  \ottnt{B} $, whose domain doesn't have the same restriction.

\begin{figure}[h]
  \drule[width=0.5\textwidth]{DT-Arrow}
  \drule[width=0.5\textwidth]{DT-AbsTm} \\\\
  \drule[width=0.5\textwidth]{DT-AppTm}
  \drule[width=0.5\textwidth]{DE-Arrow}
  \caption{Typing and definitional equality rules (floating functions)}
  \label{fig:arrow-typing}
\end{figure}

The typing rules for nondependent function types, functions, and application
are given in \cref{fig:arrow-typing}.
The domain, codomain, and entire nondependent function type are all typed at the same level.
Functions take arguments of the same level as their bodies,
and are thus applied to arguments of the same level.

This distinction between stratified dependent and unstratified nondependent functions
corresponds closely to Stratified System F:
type polymorphism is syntactically distinct from ordinary function types,
and the former forces the codomain to be a higher level while the latter doesn't.
From the perspective of Stratified System F,
the dependent types of \lang generalize stratified type polymorphism over types to include term polymorphism.

We say that the domain of these nondependent function types \emph{floats}
because unlike the stratified dependent function types,
it isn't fixed to some particular level.
The interaction between floating functions and cumulativity is where this becomes interesting.
Given a function $f$ of type $ \ottnt{A}  \mathrel{\rightarrow}  \ottnt{B} $ at level $\ottnt{j}$,
by cumulativity, it remains well typed with the same type at any level $ \ottnt{k}  \geq  \ottnt{j} $.
The level of the domain floats up from $\ottnt{j}$ to match the function at $\ottnt{k}$,
in the sense that $f$ can be applied to an argument of type $\ottnt{A}$ at any greater level $\ottnt{k}$.
This is unusual because the domain isn't contravariant
with respect to the ordering on the levels as we might expect,
and is why, as we'll see shortly, the proof of consistency in \cref{sec:semantics}
can't be straightforwardly extended to accommodate floating function types.

\section{Examples} \label{sec:examples}

\subsection{The identity function}

In the following examples, we demonstrate why floating functions are essential.
Below on the left is one way we could assign a type to the type-polymorphic identity function.
For concision, we use a pattern syntax when defining global functions
and place function arguments to the left of the definition.
(The subscript is part of the constant name.)
\begin{align*}
  & \mathsf{ id_0 } :^{ \ottsym{1} }   \Pi \mathit{ X }\!:^{ \ottsym{0} }\!  \star   \mathpunct{.}   \Pi \mathit{ \mathit{x} }\!:^{ \ottsym{0} }\!  \mathit{ X }   \mathpunct{.}   \mathit{ X }     && \mathsf{ id } :^{ \ottsym{1} }    \Pi \mathit{ X }\!:^{ \ottsym{0} }\!  \star   \mathpunct{.}   \mathit{ X }    \mathrel{\rightarrow}   \mathit{ X }    \\
  &    \mathsf{ id_0 }  \;  \mathit{ X }   \;  \mathit{ \mathit{x} }    \coloneqq   \mathit{ \mathit{x} }   &&    \mathsf{ id }  \;  \mathit{ X }   \;  \mathit{ \mathit{x} }    \coloneqq   \mathit{ \mathit{x} }  
\end{align*}

Stratification enforces that the codomain of the function type and the function body
have a higher level than that of the domain and the argument,
so the overall identity function is well typed at level 1.
While $\mathit{x}$ and $X$ have level 0 in the context of the body,
by subsumption, we can use $\mathit{x}$ at level 1 in the body as required.

Alternatively, since the return type doesn't depend on the second argument,
we can use a floating function type instead,
given above on the right.
Since we still have a dependent type quantification,
the function $  \mathit{ X }   \mathrel{\rightarrow}   \mathit{ X }  $ is still typed at level 1.
This means that $\mathit{x}$ now has level 1 directly rather than through subsumption.

So far, there's no reason to pick one over the other,
so let's look at a more involved example:
applying an identity function to itself.
This is possible due to cumulativity,
and we'll follow the corresponding Coq example below.

\begin{minted}{coq}
  Universes u0 u1.
  Constraint u0 < u1.
  Definition idid1 (id : forall (X : Type@{u1}), X -> X) :
    forall (X : Type@{u0}), X -> X :=
    id (forall (X : Type@{u0}), X -> X) (fun X => id X).
\end{minted}

Here, since \mintinline{coq}|forall (X : Type@{u0}), X -> X|
can be assigned type \mintinline{coq}|Type@{u1}|,
it can be applied as the first argument to \mintinline{coq}|id|.
While \mintinline{coq}|id| itself doesn't have this type,
we can $\eta$-expand it to a function that does,
since \mintinline{coq}|Type@{u0}| is a subtype of \mintinline{coq}|Type@{u1}|,
so \mintinline{coq}|X| can be passed to \mintinline{coq}|id|.

If we try to write the analogous definition in \sublang without using floating functions,
we find that it doesn't type check! The problematic subterm is underlined in red below.
\begin{align*}
  & \mathsf{ idid_1 } :^{ \ottsym{3} }   \Pi \mathit{ id }\!:^{ \ottsym{2} }\! \ottsym{(}   \Pi \mathit{ X }\!:^{ \ottsym{1} }\!  \star   \mathpunct{.}   \Pi \mathit{ \mathit{x} }\!:^{ \ottsym{1} }\!  \mathit{ X }   \mathpunct{.}   \mathit{ X }     \ottsym{)}  \mathpunct{.}   \Pi \mathit{ X }\!:^{ \ottsym{0} }\!  \star   \mathpunct{.}   \Pi \mathit{ \mathit{x} }\!:^{ \ottsym{0} }\!  \mathit{ X }   \mathpunct{.}   \mathit{ X }      \\
  &   \mathsf{ idid_1 }  \;  \mathit{ id }    \coloneqq     \mathit{ id }   \;  \ottsym{(}   \Pi \mathit{ X }\!:^{ \ottsym{0} }\!  \star   \mathpunct{.}   \Pi \mathit{ \mathit{x} }\!:^{ \ottsym{0} }\!  \mathit{ X }   \mathpunct{.}   \mathit{ X }     \ottsym{)}   \;   \textcolor{red}{ \underline{ \ottsym{(}     \lambda \mathit{ X } \mathpunct{.}   \lambda \mathit{ \mathit{x} } \mathpunct{.}   \mathit{ id }     \;   \mathit{ X }    \;   \mathit{ \mathit{x} }    \ottsym{)} } }   
\end{align*}

After $\eta$-expansion, $   \lambda \mathit{ X } \mathpunct{.}   \lambda \mathit{ \mathit{x} } \mathpunct{.}   \mathit{ id }     \;   \mathit{ X }    \;   \mathit{ \mathit{x} }  $ has the correct type $ \Pi \mathit{ X }\!:^{ \ottsym{0} }\!  \star   \mathpunct{.}   \Pi \mathit{ \mathit{x} }\!:^{ \ottsym{0} }\!  \mathit{ X }   \mathpunct{.}   \mathit{ X }   $,
but only at level $2$, since that's the level of $id$ itself.
Meanwhile, the second argument of $id$ expects an argument of that type but \emph{at level 1}.
We can't just raise the level annotation for that argument to 2, either,
since that would raise the level of $id$ to 3.

If we instead use floating functions for the nondependent argument,
the analogous definition then \emph{does} type check,
since the second argument of type $X$ can now be at level 2.
\begin{align*}
  & \mathsf{ idid_1 } :^{ \ottsym{2} }    \ottsym{(}    \Pi \mathit{ X }\!:^{ \ottsym{1} }\!  \star   \mathpunct{.}   \mathit{ X }    \mathrel{\rightarrow}   \mathit{ X }    \ottsym{)}  \mathrel{\rightarrow}   \Pi \mathit{ X }\!:^{ \ottsym{0} }\!  \star   \mathpunct{.}   \mathit{ X }     \mathrel{\rightarrow}   \mathit{ X }    \\
  &   \mathsf{ idid_1 }  \;  \mathit{ id }    \coloneqq     \mathit{ id }   \;  \ottsym{(}    \Pi \mathit{ X }\!:^{ \ottsym{0} }\!  \star   \mathpunct{.}   \mathit{ X }    \mathrel{\rightarrow}   \mathit{ X }    \ottsym{)}   \;  \ottsym{(}    \lambda \mathit{ X } \mathpunct{.}   \mathit{ id }    \;   \mathit{ X }    \ottsym{)}  
\end{align*}

This definition of $ \mathsf{ idid1 } $ is now pretty much shaped the same as the Coq version,
only with level annotations on domains where Coq has the corresponding level annotations on \mintinline{coq}|Type|.
If we were to turn on universe polymorphism in Coq,
it would achieve the same kind of expressivity of being able to displace $ \mathsf{ idid2 } $ in \lang.
The only difference is that while Coq merely enforces a strict inequality constraint between the levels,
in \lang the levels annotations are concrete, so even with displacement,
the distance between the two levels in the type is always $1$.

As an additional remark,
even with floating functions,
repeatedly nesting identity function self-applications is one way to non-trivially
force the level to increase.
The following definitions continue the pattern from $idid_1$,
which in the untyped setting would correspond to
$  \lambda \mathit{ id } \mathpunct{.}   \mathit{ id }    \;   \mathit{ id }  $, $   \lambda \mathit{ id } \mathpunct{.}   \mathit{ id }    \;  \ottsym{(}    \lambda \mathit{ id } \mathpunct{.}   \mathit{ id }    \;   \mathit{ id }    \ottsym{)}   \;   \mathit{ id }  $, $   \lambda \mathit{ id } \mathpunct{.}   \mathit{ id }    \;  \ottsym{(}     \lambda \mathit{ id } \mathpunct{.}   \mathit{ id }    \;  \ottsym{(}    \lambda \mathit{ id } \mathpunct{.}   \mathit{ id }    \;   \mathit{ id }    \ottsym{)}   \;   \mathit{ id }    \ottsym{)}   \;   \mathit{ id }  $,
and so on.
\begin{align*}
  & \mathsf{ idid_2 } :^{ \ottsym{3} }    \ottsym{(}    \Pi \mathit{ X }\!:^{ \ottsym{2} }\!  \star   \mathpunct{.}   \mathit{ X }    \mathrel{\rightarrow}   \mathit{ X }    \ottsym{)}  \mathrel{\rightarrow}   \Pi \mathit{ X }\!:^{ \ottsym{0} }\!  \star   \mathpunct{.}   \mathit{ X }     \mathrel{\rightarrow}   \mathit{ X }    \\
  &   \mathsf{ idid_2 }  \;  \mathit{ id }    \coloneqq      \mathit{ id }   \;  \ottsym{(}    \ottsym{(}    \Pi \mathit{ X }\!:^{ \ottsym{1} }\!  \star   \mathpunct{.}   \mathit{ X }    \mathrel{\rightarrow}   \mathit{ X }    \ottsym{)}  \mathrel{\rightarrow}   \Pi \mathit{ X }\!:^{ \ottsym{0} }\!  \star   \mathpunct{.}   \mathit{ X }     \mathrel{\rightarrow}   \mathit{ X }    \ottsym{)}   \;    \mathsf{ idid_1 }     \;  \ottsym{(}     \lambda \mathit{ X } \mathpunct{.}   \lambda \mathit{ \mathit{x} } \mathpunct{.}   \mathit{ id }     \;   \mathit{ X }    \;   \mathit{ \mathit{x} }    \ottsym{)}   \\
  & \mathsf{ idid_3 } :^{ \ottsym{4} }    \ottsym{(}    \Pi \mathit{ X }\!:^{ \ottsym{3} }\!  \star   \mathpunct{.}   \mathit{ X }    \mathrel{\rightarrow}   \mathit{ X }    \ottsym{)}  \mathrel{\rightarrow}   \Pi \mathit{ X }\!:^{ \ottsym{0} }\!  \star   \mathpunct{.}   \mathit{ X }     \mathrel{\rightarrow}   \mathit{ X }    \\
  &   \mathsf{ idid_3 }  \;  \mathit{ id }    \coloneqq      \mathit{ id }   \;  \ottsym{(}    \ottsym{(}    \Pi \mathit{ X }\!:^{ \ottsym{2} }\!  \star   \mathpunct{.}   \mathit{ X }    \mathrel{\rightarrow}   \mathit{ X }    \ottsym{)}  \mathrel{\rightarrow}   \Pi \mathit{ X }\!:^{ \ottsym{0} }\!  \star   \mathpunct{.}   \mathit{ X }     \mathrel{\rightarrow}   \mathit{ X }    \ottsym{)}   \;    \mathsf{ idid_2 }     \;  \ottsym{(}     \lambda \mathit{ X } \mathpunct{.}   \lambda \mathit{ \mathit{x} } \mathpunct{.}   \mathit{ id }     \;   \mathit{ X }    \;   \mathit{ \mathit{x} }    \ottsym{)}  
\end{align*}

All of $  \mathit{ idid_1 }   \;  \ottsym{(}   \lambda \mathit{ X } \mathpunct{.}   \lambda \mathit{ \mathit{x} } \mathpunct{.}   \mathit{ \mathit{x} }     \ottsym{)} $, $  \mathit{ idid_2 }   \;  \ottsym{(}   \lambda \mathit{ X } \mathpunct{.}   \lambda \mathit{ \mathit{x} } \mathpunct{.}   \mathit{ \mathit{x} }     \ottsym{)} $, and $  \mathit{ idid_3 }   \;  \ottsym{(}   \lambda \mathit{ X } \mathpunct{.}   \lambda \mathit{ \mathit{x} } \mathpunct{.}   \mathit{ \mathit{x} }     \ottsym{)} $
reduce to $ \lambda \mathit{ X } \mathpunct{.}   \lambda \mathit{ \mathit{x} } \mathpunct{.}   \mathit{ \mathit{x} }   $.

\subsection{Decidable types}

The following example demonstrates a more substantial use of \lang
in the form of type constructors as floating functions and how they interact with cumulativity.
Later in \cref{sec:implementation} we'll consider datatypes with parameters, but for now,
consider the following Church encoding \citep{encoding} of decidable types,
which additionally uses negation defined as implication into the empty type.
\begin{align*}
  & \mathsf{ neg } :^{ \ottsym{0} }    \star   \mathrel{\rightarrow}   \star    && \mathsf{ yes } :^{ \ottsym{1} }     \Pi \mathit{ X }\!:^{ \ottsym{0} }\!  \star   \mathpunct{.}   \mathit{ X }    \mathrel{\rightarrow}   \mathsf{ Dec }    \;   \mathit{ X }    \\
  &   \mathsf{ neg }  \;  \mathit{ X }    \coloneqq    \mathit{ X }   \mathrel{\rightarrow}   \bot         &&    \mathsf{ yes }  \;  \mathit{ X }   \;  \mathit{ \mathit{x} }    \coloneqq    \lambda \mathit{ Z } \mathpunct{.}   \lambda \mathit{ f } \mathpunct{.}   \lambda \mathit{ g } \mathpunct{.}   \mathit{ f }      \;   \mathit{ \mathit{x} }    \\
  & \mathsf{ Dec } :^{ \ottsym{1} }    \star   \mathrel{\rightarrow}   \star    && \mathsf{ no } :^{ \ottsym{1} }      \Pi \mathit{ X }\!:^{ \ottsym{0} }\!  \star   \mathpunct{.}   \mathsf{ neg }    \;   \mathit{ X }    \mathrel{\rightarrow}   \mathsf{ Dec }    \;   \mathit{ X }    \\
  &   \mathsf{ Dec }  \;  \mathit{ X }    \coloneqq     \Pi \mathit{ Z }\!:^{ \ottsym{0} }\!  \star   \mathpunct{.}  \ottsym{(}    \mathit{ X }   \mathrel{\rightarrow}   \mathit{ Z }    \ottsym{)}   \mathrel{\rightarrow}  \ottsym{(}     \mathsf{ neg }   \;   \mathit{ X }    \mathrel{\rightarrow}   \mathit{ Z }    \ottsym{)}   \mathrel{\rightarrow}   \mathit{ Z }    &&    \mathsf{ no }  \;  \mathit{ X }   \;  \mathit{ nx }    \coloneqq    \lambda \mathit{ Z } \mathpunct{.}   \lambda \mathit{ f } \mathpunct{.}   \lambda \mathit{ g } \mathpunct{.}   \mathit{ g }      \;   \mathit{ nx }   
\end{align*}

The $  \mathsf{ yes }   \;   \mathit{ X }  $ constructor decides $X$ by a witness,
while the $  \mathsf{ no }   \;   \mathit{ X }  $ constructor decides $X$ by its refutation.
We're able to show that deciding a given type is irrefutable\punctstack{.}%
\footnote{Note this differs from irrefutability of the law of excluded middle,
$  \mathsf{ neg }   \;  \ottsym{(}    \mathsf{ neg }   \;  \ottsym{(}    \Pi \mathit{ X }\!:^{ \ottsym{0} }\!  \star   \mathpunct{.}   \mathsf{ Dec }    \;   \mathit{ X }    \ottsym{)}   \ottsym{)} $,
which cannot be proven constructively.}
\begin{align*}
  & \mathsf{ irrDec } :    \Pi \mathit{ X }\!:^{ \ottsym{0} }\!  \star   \mathpunct{.}   \mathsf{ neg }    \;  \ottsym{(}    \mathsf{ neg }   \;  \ottsym{(}    \mathsf{ Dec }   \;   \mathit{ X }    \ottsym{)}   \ottsym{)}   \\
  &    \mathsf{ irrDec }  \;  \mathit{ X }   \;  \mathit{ ndec }    \coloneqq    \mathit{ ndec }   \;  \ottsym{(}     \mathsf{ no }   \;   \mathit{ X }    \;  \ottsym{(}    \lambda \mathit{ \mathit{x} } \mathpunct{.}   \mathit{ ndec }    \;  \ottsym{(}     \mathsf{ yes }   \;   \mathit{ X }    \;   \mathit{ \mathit{x} }    \ottsym{)}   \ottsym{)}   \ottsym{)}  
\end{align*}

The same exercise of trying to define $ \mathsf{ neg } $ and $ \mathsf{ Dec } $
using only dependent functions and not floating functions
to the same effect of no longer being able to type check $ \mathsf{ irrDec } $,
even if we allow ourselves to use displacement.
More interestingly, let's now compare these definitions
to the corresponding ones in Agda.

\begin{minted}{agda}
  {-# OPTIONS --cumulativity #-}
  open import Agda.Primitive using (lzero ; lsuc)
  open import Data.Empty using (⊥)
  neg : ∀ ℓ → Set ℓ → Set ℓ
  neg ℓ X = X → ⊥
  Dec : ∀ ℓ → Set (lsuc ℓ) → Set (lsuc ℓ)
  Dec ℓ X = (Z : Set ℓ) → (X → Z) → (neg (lsuc ℓ) X → Z) → Z
  yes : ∀ ℓ (X : Set ℓ) → X → Dec ℓ X
  yes ℓ X x = λ Z f g → f x
  no : ∀ ℓ (X : Set ℓ) → neg ℓ X → Dec ℓ X
  no ℓ X nx = λ Z f g → g nx
\end{minted}

They must all be universe polymorphic to capture the expressivity of floating functions.
For instance, to talk about the negation of a type at level 1,
by cumulativity it suffices to use $ \mathsf{ neg } $ (without displacement!) in \lang,
but we must use \mintinline{agda}|neg (lsuc lzero)| in Agda.
Effectively, the \lang type $  \star   \mathrel{\rightarrow}   \star  $ represents not merely \mintinline{agda}|Set → Set|
but, by cumulativity, all types \mintinline{agda}|Set ℓ → Set ℓ| for every \mintinline{agda}|ℓ|.

\subsection{Leibniz equality}

Although nondependent functions can often benefit from a floating domain,
sometimes we don't want the domain to float.
Here, we turn to a simple application of dependent types with Leibniz equality \citep{leibniz, mltt-type-in-type}
to demonstrate a situation where the level of the domain needs to be fixed
to something strictly smaller than that of the codomain
even when the codomain doesn't depend on the function argument.
\begin{align*}
  & \mathsf{ eq } :^{ \ottsym{1} }     \Pi \mathit{ X }\!:^{ \ottsym{0} }\!  \star   \mathpunct{.}   \mathit{ X }    \mathrel{\rightarrow}   \mathit{ X }    \mathrel{\rightarrow}   \star    && \mathsf{ refl } :^{ \ottsym{1} }      \Pi \mathit{ X }\!:^{ \ottsym{0} }\!  \star   \mathpunct{.}   \Pi \mathit{ \mathit{x} }\!:^{ \ottsym{0} }\!  \mathit{ X }   \mathpunct{.}   \mathsf{ eq }     \;   \mathit{ X }    \;   \mathit{ \mathit{x} }    \;   \mathit{ \mathit{x} }    \\
  &     \mathsf{ eq }  \;  \mathit{ X }   \;  \mathit{ \mathit{x} }   \;  \mathit{ \mathit{y} }    \coloneqq      \Pi \mathit{ P }\!:^{ \ottsym{0} }\!   \mathit{ X }   \mathrel{\rightarrow}   \star    \mathpunct{.}   \mathit{ P }    \;   \mathit{ \mathit{x} }    \mathrel{\rightarrow}   \mathit{ P }    \;   \mathit{ \mathit{y} }    &&      \mathsf{ refl }  \;  \mathit{ X }   \;  \mathit{ \mathit{x} }   \;  \mathit{ P }   \;  \mathit{ px }    \coloneqq   \mathit{ px }  
\end{align*}

An equality $    \mathsf{ eq }   \;  \ottnt{A}   \;  \ottnt{a}   \;  \ottnt{b} $ states that two terms are equal
if given any predicate $P$, a proof of $  \mathit{ P }   \;  \ottnt{a} $ yields a proof of $  \mathit{ P }   \;  \ottnt{b} $;
in other words, $\ottnt{a}$ and $\ottnt{b}$ are indiscernible.
The proof of reflexivity of Leibniz equality should be unsurprising.

We might try to define a predicate stating that a given type $X$ is a mere proposition,
\ie that all of its inhabitants are equal,
and give it a nondependent function type.
\begin{align*}
  & \mathsf{ isProp } :^{ \ottsym{0} }    \star   \mathrel{\rightarrow}   \star    \\
  &   \mathsf{ isProp }  \;  \mathit{ X }    \coloneqq   \textcolor{red}{ \underline{     \Pi \mathit{ \mathit{x} }\!:^{ \ottsym{0} }\!  \mathit{ X }   \mathpunct{.}   \Pi \mathit{ \mathit{y} }\!:^{ \ottsym{0} }\!  \mathit{ X }   \mathpunct{.}   \mathsf{ eq }     \;   \mathit{ X }    \;   \mathit{ \mathit{x} }    \;   \mathit{ \mathit{y} }   } }  
\end{align*}

But this doesn't type check, since the body contains an equality over elements of $X$,
which necessarily has level 1 rather than the expected level 0.
We must assign $ \mathsf{ isProp } $ a stratified function type, given below on the left;
informally, stratification propagates dependency information not only from the codomain,
but also from the function body.
\begin{align*}
  & \mathsf{ isProp } :^{ \ottsym{1} }   \Pi \mathit{ X }\!:^{ \ottsym{0} }\!  \star   \mathpunct{.}   \star    && \mathsf{ isSet } :^{ \ottsym{2} }   \Pi \mathit{ X }\!:^{ \ottsym{0} }\!  \star   \mathpunct{.}   \star    \\
  &   \mathsf{ isProp }  \;  \mathit{ X }    \coloneqq      \Pi \mathit{ \mathit{x} }\!:^{ \ottsym{0} }\!  \mathit{ X }   \mathpunct{.}   \Pi \mathit{ \mathit{y} }\!:^{ \ottsym{0} }\!  \mathit{ X }   \mathpunct{.}   \mathsf{ eq }     \;   \mathit{ X }    \;   \mathit{ \mathit{x} }    \;   \mathit{ \mathit{y} }    &&   \mathsf{ isSet }  \;  \mathit{ X }    \coloneqq    \Pi \mathit{ \mathit{x} }\!:^{ \ottsym{0} }\!  \mathit{ X }   \mathpunct{.}   \Pi \mathit{ \mathit{y} }\!:^{ \ottsym{0} }\!  \mathit{ X }   \mathpunct{.}   \mathsf{ isProp }^{ \ottsym{1} }     \;  \ottsym{(}      \mathsf{ eq }   \;   \mathit{ X }    \;   \mathit{ \mathit{x} }    \;   \mathit{ \mathit{y} }    \ottsym{)}  
\end{align*}

Going one further, we define above on the right a predicate $ \mathsf{ isSet } $
stating that $X$ is an h-set~\citep{hottbook},
or that its equalities are mere propositions,
by using a displaced $ \mathsf{ isProp } $ so that we can reuse the definition at a higher level;
here, $ \mathsf{ isProp }^{ \ottsym{1} } $ now has type $ \Pi \mathit{ X }\!:^{ \ottsym{1} }\!  \star   \mathpunct{.}   \star  $ at level 2.
Once again, despite the type of $ \mathsf{ isSet } $ not being an actual dependent function type,
here we need to fix the level of the domain.

\section{Metatheory} \label{sec:metatheory}

\subsection{Consistency of \sublang} \label{sec:semantics}

We use Agda to mechanize a proof of logical consistency ---
that no closed inhabitant of the empty type exists ---
for \sublang, which excludes floating nondependent functions.
For simplicity, the mechanization also excludes global definitions and displaced constants,
which shouldn't affect consistency: if there is a closed inhabitant of the empty type that uses global definitions,
then there is a closed inhabitant of the empty type under the empty signature
by inlining all global definitions.
The proof files are available
\ifanonymous in the supplementary materials
\else at \publicrepo{}
\fi
under the \texttt{agda/} directory.
The only axiom we use is function extensionality\punctstack{.}%
\footnote{\footfilethm{agda/accessibility.agda}{funext,funext'}}
\jcxz{TODO [low priority]: Maybe we should prove in the Coq proofs that if $ \Delta ;  \varnothing  \vdash  \ottnt{a}  :^{ \ottnt{k} }  \ottnt{A} $
then there exist $ \Delta  \vdash  \ottnt{a}  \leadsto^{\ast}  \ottnt{a'} $ and $ \Delta  \vdash  \ottnt{A}  \leadsto^{\ast}  \ottnt{A'} $
such that $ \varnothing ;  \varnothing  \vdash  \ottnt{a'}  :^{ \ottnt{k} }  \ottnt{A'} $,
where all $\delta$-reductions have been performed.}

The core construction of the consistency proof is a three-place logical relation
\fbox{$\ottnt{a} \in \interp{\ottnt{A}}_{\ottnt{k}}$} among a term, its type, and its level,
which we would aspirationally like to define as follows,
using $\mathbf{0}$ for falsehood, $\mathbf{1}$ for truthhood,
$\wedge$ for conjunction, $\longrightarrow$ for implication,
and $\forall$ and $\exists$ for universal and existential quantification
in our working metatheory.
\begin{align*}
   \star  \in \interp{ \star }_{k}
    &\triangleq \mathbf{1}
    & \Pi \mathit{ \mathit{x} }\!:^{ \ottnt{j} }\! \ottnt{A}  \mathpunct{.}  \ottnt{B}  \in \interp{ \star }_{k}
    &\triangleq j < k \wedge A \in \interp{ \star }_{j}
      \wedge (\forall y \mathpunct{.} y \in \interp{\ottnt{A}}_{j} \longrightarrow  \ottnt{B}  \mathopen{ \{ }   \mathit{ \mathit{y} }   / \mathit{ \mathit{x} } \mathclose{ \} }  \in \interp{ \star }_{k}) \\
   \bot  \in \interp{ \star }_{k}
    &\triangleq \mathbf{1}
    &f \in \interp{ \Pi \mathit{ \mathit{x} }\!:^{ \ottnt{j} }\! \ottnt{A}  \mathpunct{.}  \ottnt{B} }_{k}
    &\triangleq \forall y \mathpunct{.} y \in \interp{\ottnt{A}}_{j} \longrightarrow   \mathit{ f }   \;   \mathit{ \mathit{y} }   \in \interp{ \ottnt{B}  \mathopen{ \{ }   \mathit{ \mathit{y} }   / \mathit{ \mathit{x} } \mathclose{ \} } }_{k} \\
  \ottnt{a} \in \interp{ \bot }_{k}
    &\triangleq \mathbf{0}
    &\ottnt{a} \in \interp{\ottnt{A}}_{k}
    &\triangleq \exists B \mathpunct{.}  \ottnt{A}   \equiv   \ottnt{B}  \wedge \ottnt{a} \in \interp{\ottnt{B}}_{k}
\end{align*}

However, this definition isn't necessarily well formed.
It isn't defined recursively on the structure of the terms or the types,
because in the cases involving dependent functions,
we need to talk about the substitution $ \ottnt{B}  \mathopen{ \{ }   \mathit{ \mathit{y} }   / \mathit{ \mathit{x} } \mathclose{ \} } $.
It isn't defined inductively, either,
because again in the dependent function case,
the inductive itself appears to the left of an implication as $y \in \interp{\ottnt{A}}_j$,
making the inductive definition non-strictly-positive.

The solution is to define the logical relation as an inductive--recursive definition~\cite{IR}.
This design is adapted from a concise proof of consistency for MLTT in Coq by Liu~\cite{yiyun},
which uses an impredicative encoding in place of induction--recursion.
This is a simplified and pared down adaptation of a proof of decidability of conversion for MLTT in Coq
by Adjedj, Lennon-Bertrand, Maillard, P\'edrot, and Pujet~\cite{MLalaCoq},
which in turn uses a predicative encoding to adapt a proof of decidability of conversion for MLTT in Agda
by Abel, \"Ohman, and Vezzosi~\cite{conversion}
that uses induction--recursion.

Below is a sketch of the inductive--recursive definition,
which splits the logical relation into two parts:
an inductive predicate on types and their levels \fbox{$\interp{\ottnt{A}}_{k}$},
and relation between types and terms defined recursively on the predicate on the type,
which we continue to write as \fbox{$a \in \interp{\ottnt{A}}_k$}.
\begin{mathpar}
  \inferrule{~}{\interp{ \star }_k} \and
  \inferrule{~}{\interp{ \bot }_k} \and
  \inferrule{
    j < k \and
    \interp{\ottnt{A}}_{j} \and
    \forall y \mathpunct{.} y \in \interp{\ottnt{A}}_{j} \longrightarrow \interp{ \ottnt{B}  \mathopen{ \{ }   \mathit{ \mathit{y} }   / \mathit{ \mathit{x} } \mathclose{ \} } }_{k}
  }{\interp{ \Pi \mathit{ \mathit{x} }\!:^{ \ottnt{j} }\! \ottnt{A}  \mathpunct{.}  \ottnt{B} }_k} \and
  \inferrule{
     \ottnt{A}  \Rightarrow  \ottnt{B}  \and
    \interp{\ottnt{B}}_k
  }{\interp{\ottnt{A}}_k}
\end{mathpar}
\begin{align*}
  A \in \interp{ \star }_k
    &\triangleq \interp{\ottnt{A}}_k
    &f \in \interp{{ \Pi \mathit{ \mathit{x} }\!:^{ \ottnt{j} }\! \ottnt{A}  \mathpunct{.}  \ottnt{B} }}_k
    &\triangleq \forall y \mathpunct{.} y \in \interp{\ottnt{A}}_{j} \longrightarrow   \mathit{ f }   \;   \mathit{ \mathit{y} }   \in \interp{ \ottnt{B}  \mathopen{ \{ }   \mathit{ \mathit{y} }   / \mathit{ \mathit{x} } \mathclose{ \} } }_k \\
  a \in \interp{ \bot }_k
    &\triangleq \mathbf{0}
    &a \in \interp{{\ottnt{A}}}_k
    &\triangleq a \in \interp{\ottnt{B}}_k \quad (\textit{where} ~  \ottnt{A}  \Rightarrow  \ottnt{B} )
\end{align*}

In the last inductive rule, in place of $ \ottnt{A}   \equiv   \ottnt{B} $,
we instead use parallel reduction \fbox{$ \ottnt{A}  \Rightarrow  \ottnt{B} $},
which is a reduction relation describing all visible reductions being performed in parallel from the inside out.
This is justified by the following lemma,
where \fbox{$ \ottnt{A}  \Rightarrow^{\ast}  \ottnt{B} $} is the reflexive, transitive closure of $ \ottnt{A}  \Rightarrow  \ottnt{B} $.

\begin{lemma}[Implementation of definitional equality]\stack{. }\footnote{\footfilethm{agda/typing.agda}{≈-⇔}}
  $ \ottnt{A}   \equiv   \ottnt{B} $ iff there exists some $\ottnt{C}$ such that $ \ottnt{A}  \mathrel{ \Rightarrow^{\ast} }  \ottnt{C}  \mathrel{ \prescript{\ast}{}{\Leftarrow} }  \ottnt{B} $,
  which we write as \fbox{$ \ottnt{A}  \Leftrightarrow  \ottnt{B} $}.
\end{lemma}

Even now, this inductive--recursive definition is \emph{still} not well formed.
In particular, in the inductive rule for dependent functions,
if $\ottnt{A}$ is $ \star $, then by the recursive case for the universe,
$\interp{\mathit{y}}_j$ could again appear to the left of an implication.
However, we know that $j < k$, which we can exploit to stratify the logical relation
just as we stratify typing judgements.
We do so by parametrizing each logical relation at level $k$
by an abstract logical relation defined at all strictly lower levels $j < k$,
then at the end tying the knot by instantiating them via \wellfounded induction on levels.
This technique is adapted from an Agda model of a universe hierarchy by Kov\'acs~\cite{universes},
which originates from McBride's redundancy-free construction of a universe hierarchy~\cite[Section~6.3.1]{datadata}.
As the constructions are now fairly involved,
we defer to the proof file\footnote{\footfile{agda/semantics.agda}}
for the full definitions,
in particular \mintinline{agda}|U| for the inductive predicate
and \mintinline{agda}|el| for the recursive relation.
For the purposes of exposition, we continue to use the old notation.

Because the logical relation only handles closed terms,
we deal with contexts and simultaneous substitutions $\sigma$ separately
by relating the two via yet another inductive--recursive definition,
with a predicate on contexts \fbox{$\interp{\Gamma}$}
and a relation between substitutions and contexts \fbox{$\sigma \in \interp{\Gamma}$}.
Here, $ \ottnt{A} \{ \sigma \} $ denotes applying the substitution $\sigma$ to the term $\ottnt{A}$,
and $ \sigma [  \mathit{x}  ] $ denotes the term which $\sigma$ substitutes for $\mathit{x}$%
\punctstack{.}\footnote{The mechanization uses de Bruijn indexing;
various index-shifting operations on substitutions are omitted for concision.}
\vspace{-2.5\baselineskip}
\begin{multicols}{2}
\begin{mathpar}
  \inferrule{~}{\interp{ \varnothing }}
  \quad
  \inferrule{
    \interp{\Gamma} \and
    \forall \sigma \mathpunct{.} \sigma \in \interp{\Gamma} \longrightarrow \interp{ \ottnt{A} \{ \sigma \} }_{k}
  }{\interp{ \Gamma ,   \mathit{x}    :^{ \ottnt{k} }  \ottnt{A}   }}
\end{mathpar}

\begin{align*}
  \sigma \in \interp{ \varnothing } &\triangleq \mathbf{1} \\
  \sigma \in \interp{ \Gamma ,   \mathit{x}    :^{ \ottnt{k} }  \ottnt{A}   } &\triangleq \sigma \in \interp{\Gamma} \wedge  \sigma [  \mathit{x}  ]  \in \interp{ \ottnt{A} \{ \sigma \} }_{k}
\end{align*}
\end{multicols}

The most important lemmas that are needed are semantic cumulativity,
semantic conversion, and backward preservation.

\begin{lemma}[Cumulativity]\stack{. }\footnote{\footfilethm{agda/semantics.agda}{cumU,cumEl}}
  If $j < k$ and $\interp{\ottnt{A}}_{j}$ then $\interp{\ottnt{A}}_{k}$,
  and if $a \in \interp{\ottnt{A}}_{j}$ then $a \in \interp{\ottnt{A}}_{k}$.
\end{lemma}

\begin{lemma}[Conversion]\stack{. }\footnote{\footfilethm{agda/semantics.agda}{⇔-U,⇔-el}}
  If $ \ottnt{A}  \Leftrightarrow  \ottnt{B} $ and $\interp{\ottnt{A}}_{k}$ then $\interp{\ottnt{B}}_{k}$,
  and if $a \in \interp{\ottnt{A}}_{k}$ then $a \in \interp{\ottnt{B}}_{k}$.
\end{lemma}

\begin{lemma}[Backward preservation]\stack{. }\footnote{\footfilethm{agda/semantics.agda}{⇒⋆-el}}
  If $ \ottnt{a}  \Rightarrow^{\ast}  \ottnt{b} $ and $b \in \interp{\ottnt{A}}_{k}$ then $a \in \interp{\ottnt{A}}_{k}$.
\end{lemma}

We can now prove the fundamental theorem of soundness
of typing judgements with respect to the logical relation
by induction on typing derivations,
and consistency follows as a corollary.

\begin{theorem}[Soundness]\stack{. }\footnote{\footfilethm{agda/soundness.agda}{soundness}}
  Suppose $\interp{\Gamma}$ and $\sigma \in \interp{\Gamma}$.
  If $ \Gamma  \vdash  \ottnt{a}  :^{ \ottnt{k} }  \ottnt{A} $, then $\interp{ \ottnt{A} \{ \sigma \} }_{k}$ and $ \ottnt{a} \{ \sigma \}  \in \interp{ \ottnt{A} \{ \sigma \} }_{k}$.
\end{theorem}

\begin{corollary}[Consistency]\stack{. }\footnote{\footfilethm{agda/consistency.agda}{consistency}}
  There are no $\ottnt{b}$, $k$ such that $ \varnothing  \vdash  \ottnt{b}  :^{ \ottnt{k} }   \bot  $.
\end{corollary}

\subsubsection{The problem with floating functions} \label{sec:proof-attempt}

This proof can't be extended to the full \lang.
While floating nondependent function types can be added to the logical relation directly as below,
cumulativity will no longer hold.
\begin{mathpar}
  \inferrule{\interp{\ottnt{A}}_{k} \and \interp{\ottnt{B}}_{k}}{\interp{ \ottnt{A}  \mathrel{\rightarrow}  \ottnt{B} }_{k}}
  \and
  {f \in \interp{ \ottnt{A}  \mathrel{\rightarrow}  \ottnt{B} }_{k} \triangleq \forall x \mathpunct{.} x \in \interp{\ottnt{A}}_{k} \longrightarrow   \mathit{ f }   \;   \mathit{ \mathit{x} }   \in \interp{\ottnt{B}}_{k}}
\end{mathpar}

In particular, given $f \in \interp{ \ottnt{A}  \mathrel{\rightarrow}  \ottnt{B} }_{j}$, when trying to show $f \in \interp{ \ottnt{A}  \mathrel{\rightarrow}  \ottnt{B} }_{k}$,
we have by definition $\forall x \mathpunct{.} x \in \interp{\ottnt{A}}_{j} \longrightarrow   \mathit{ f }   \;   \mathit{ \mathit{x} }   \in \interp{\ottnt{B}}_{j}$,
a term $x$, and $x \in \interp{\ottnt{A}}_{k}$, but no way to cast the latter into $x \in \interp{\ottnt{A}}_{j}$
to obtain $  \mathit{ f }   \;   \mathit{ \mathit{x} }   \in \interp{\ottnt{B}}_{k}$ as desired via the induction hypothesis,
because such a cast would go \emph{downwards} from a higher level $k$ to a lower level $j$,
rather than the other way around as provided by the induction hypothesis.
Trying to incorporate the desired property into the relation, perhaps by defining it as
$\forall \ell \ge k \mathpunct{.} \forall x \mathpunct{.} x \in \interp{\ottnt{A}}_{\ell} \longrightarrow   \mathit{ f }   \;   \mathit{ \mathit{x} }   \in \interp{\ottnt{B}}_{k}$,
would break the careful stratification of the logical relation that we've set up.

\subsection{Type safety of \lang} \label{sec:syntactics}

While we haven't yet proven its consistency,
we have proven type safety of the full \lang.
We use Coq to mechanize the syntactic metatheory of
the typing, context formation, and signature formation judgements of \lang,
recalling that this covers all of stratified dependent functions,
floating nondependent functions,
and displaced constants.
We also use Ott~\citep{Ott} along with the Coq tools LNgen~\citep{LNgen} and Metalib~\citep{metalib}
to represent syntax and judgements
and to handle their locally-nameless representation in Coq.
The proof scripts are available
\ifanonymous in the supplementary materials
\else at \publicrepo{}
\fi
under the \texttt{coq/} directory.

We begin with some basic common properties of type systems,
namely weakening, substitution, and regularity lemmas,
as well as a generalized displaceability lemma that's simple to show.
Next, we introduce a notion of \emph{restriction},
which formalizes the idea that lower judgements can't depend on higher ones,
along with a notion of \emph{restricted floating},
which is crucial for proving that floating function types are \emph{syntactically} cumulative.
Only then are we able to prove type safety.

As we haven't mechanized the syntactic metatheory of definitional equality $ \Delta  \vdash  \ottnt{A} \equiv \ottnt{B} $,
we state as axioms some standard, provable properties~\cite[Section 5.2]{barendregt},
which are orthogonal to stratification
and only used in the final proof of type safety.
The equivalent lemmas for \sublang, however, have been mechanized in Agda%
\footnote{\footfile{agda/reduction.agda}} as part of the consistency proof.

\begin{axiom}[Function type injectivity]\stack{. }\footnote{\footfilethm{coq/axioms.v}{DEquiv\_Arrow\_inj1,DEquiv\_Arrow\_inj2,DEquiv\_Pi\_inj1,DEquiv\_Pi\_inj2}}
  If $ \Delta  \vdash   \ottnt{A_{{\mathrm{1}}}}  \mathrel{\rightarrow}  \ottnt{B_{{\mathrm{1}}}}  \equiv  \ottnt{A_{{\mathrm{2}}}}  \mathrel{\rightarrow}  \ottnt{B_{{\mathrm{2}}}}  $ then $ \Delta  \vdash  \ottnt{A_{{\mathrm{1}}}} \equiv \ottnt{A_{{\mathrm{2}}}} $ and $ \Delta  \vdash  \ottnt{B_{{\mathrm{1}}}} \equiv \ottnt{B_{{\mathrm{2}}}} $;
  if $  \Pi \mathit{ \mathit{x} }\!:^{ \ottnt{j_{{\mathrm{1}}}} }\! \ottnt{A_{{\mathrm{1}}}}  \mathpunct{.}  \ottnt{B_{{\mathrm{1}}}}    \equiv    \Pi \mathit{ \mathit{x} }\!:^{ \ottnt{j_{{\mathrm{2}}}} }\! \ottnt{A_{{\mathrm{2}}}}  \mathpunct{.}  \ottnt{B_{{\mathrm{2}}}}  $
  then $ \Delta  \vdash  \ottnt{A_{{\mathrm{1}}}} \equiv \ottnt{A_{{\mathrm{2}}}} $, $\ottnt{j_{{\mathrm{1}}}}  \ottsym{=}  \ottnt{j_{{\mathrm{2}}}}$, and $ \Delta  \vdash  \ottnt{B_{{\mathrm{1}}}} \equiv \ottnt{B_{{\mathrm{2}}}} $.
\end{axiom}

\begin{axiom}[Consistency of definitional equality]\stack{. }\footnote{\footfilethm{coq/axioms.v}{ineq\_*}}
  If $ \Delta  \vdash  \ottnt{A} \equiv \ottnt{B} $ then $\ottnt{A}$ and $\ottnt{B}$ do not have different head forms.
\end{axiom}

\subsubsection{Basic properties}

We can extend the ordering between levels $\ottnt{j}  \leq  \ottnt{k}$ to an ordering
between contexts \fbox{$ \Gamma_{{\mathrm{1}}}   \leq   \Gamma_{{\mathrm{2}}} $}; that is,
if $\ottnt{j}  \leq  \ottnt{k}$, then $   \Gamma ,   \mathit{x}    :^{ \ottnt{j} }  \ottnt{A}       \leq     \Gamma ,   \mathit{x}    :^{ \ottnt{k} }  \ottnt{A}     $.
At the same time, we also incorporate the
idea of weakening into this relation, so $   \Gamma ,   \mathit{x}    :^{ \ottnt{k} }  \ottnt{A}       \leq   \Gamma $.
Stronger contexts have higher levels and fewer assumptions.
This ordering is contravariant in the typing judgement: we can lower the
context without destroying typeability. This result subsumes a standard
weakening lemma.

\begin{lemma}[Weakening]\stack{. }\footnote{\footfilethm{coq/ctx.v}{DTyping\_SubG}}
If $ \Delta ;  \Gamma  \vdash  \ottnt{a}  :^{ \ottnt{k} }  \ottnt{A} $ and $\Delta  \vdash  \Gamma'$ and $ \Gamma'   \leq   \Gamma $ then $ \Delta ;  \Gamma'  \vdash  \ottnt{a}  :^{ \ottnt{k} }  \ottnt{A} $.
\end{lemma}

The substitution lemma reflects the idea that an assumption $ \mathit{x}    :^{ \ottnt{k} }  \ottnt{B}  $ is a
hypothetical judgement. The variable $\mathit{x}$ stands for any typing derivation
of the appropriate type and level.

\begin{lemma}[Substitution]\stack{. }\footnote{\footfilethm{coq/subst.v}{DTyping\_subst}}
If $ \Delta ;     \Gamma_{{\mathrm{1}}} ,   \mathit{x}    :^{ \ottnt{j} }  \ottnt{B}     ,  \Gamma_{{\mathrm{2}}}   \vdash  \ottnt{a}  :^{ \ottnt{k} }  \ottnt{A} $ and $ \Delta ;  \Gamma_{{\mathrm{1}}}  \vdash  \ottnt{b}  :^{ \ottnt{j} }  \ottnt{B} $ then 
  $ \Delta ;   \Gamma_{{\mathrm{1}}} ,  \Gamma_{{\mathrm{2}}}   \ottsym{\{}  \ottnt{b}  \ottsym{/}  \mathit{x}  \ottsym{\}}  \vdash   \ottnt{a}  \mathopen{ \{ }  \ottnt{b}  / \mathit{ \mathit{x} } \mathclose{ \} }   :^{ \ottnt{k} }   \ottnt{A}  \mathopen{ \{ }  \ottnt{b}  / \mathit{ \mathit{x} } \mathclose{ \} }  $.
\end{lemma}

Typing judgements themselves ensure the \wellformedness of their components;
in particular, if a term type checks, then its type can be typed at the same level.
Because our type system includes the non--syntax-directed \rref{T-Conv},
the proof of this lemma depends on several inversion lemmas, omitted here.

\begin{lemma}[Regularity]\stack{. }\footnote{\footfilethm{coq/ctx.v}{DCtx\_DSig}, \footfilethm{coq/inversion.v}{DTyping\_DCtx}, \footfilethm{coq/ctx.v}{DTyping\_regularity}} \label{lem:regularity}
If $ \Delta ;  \Gamma  \vdash  \ottnt{a}  :^{ \ottnt{k} }  \ottnt{A} $ then $\vdash  \Delta$ and $\Delta  \vdash  \Gamma$ and $ \Delta ;  \Gamma  \vdash  \ottnt{A}  :^{ \ottnt{k} }   \star  $
\end{lemma}

Generalizing displaceability in an empty context,
derivations can be displaced wholesale by also incrementing contexts,
written $ \Gamma ^{+ \ottnt{i} } $, where $ \ottsym{(}   \Gamma ,   \mathit{x}    :^{ \ottnt{k} }  \ottnt{A}     \ottsym{)} ^{+ \ottnt{i} }  =   \Gamma ^{+ \ottnt{i} }  ,   \mathit{x}    :^{  \ottnt{k}  \ottsym{+}  \ottnt{i}  }   \ottnt{A} ^{+ \ottnt{i} }    $.

\begin{lemma}[Displaceability]\stack{. }\footnote{\footfilethm{coq/ctx.v}{DTyping\_incr}} \label{lem:incr}
If $ \Delta ;  \Gamma  \vdash  \ottnt{a}  :^{ \ottnt{k} }  \ottnt{A} $ then $ \Delta ;   \Gamma ^{+ \ottnt{j} }   \vdash   \ottnt{a} ^{+ \ottnt{j} }   :^{ \ottnt{j}  \ottsym{+}  \ottnt{k} }   \ottnt{A} ^{+ \ottnt{j} }  $.
\end{lemma}

If we displace a context, the result might not be stronger
because displacement may modify the types in the assumptions.
In other words, it is \emph{not} the case that $ \Gamma   \leq    \Gamma ^{+ \ottnt{k} }  $.

\subsection{Restriction}

The key idea of stratification is that a judgement at level $\ottnt{k}$
is only allowed to depend on judgements at the same or lower levels.
One way to observe this property is through a form of strengthening result,
which allows variables from higher levels to be removed from the
context and contexts to be truncated at any level.
Formally, we define the \emph{restriction} operation,
written $ \lceil \Gamma \rceil^{ \ottnt{k} } $,
which filters out all assumptions from the context with level greater than $k$.
A restricted context can be stronger since it could contain fewer assumptions.

\begin{lemma}[Restriction]\stack{. }\footnote{\footfilethm{coq/ctx.v}{DSig\_DCtx\_DTyping\_restriction}} \label{lem:restriction}
If $\Delta  \vdash  \Gamma$ then $\Delta  \vdash   \lceil \Gamma \rceil^{ \ottnt{k} } $ for any $\ottnt{k}$,
and if $ \Delta ;  \Gamma  \vdash  \ottnt{a}  :^{ \ottnt{k} }  \ottnt{A} $ then $ \Delta ;   \lceil \Gamma \rceil^{ \ottnt{k} }   \vdash  \ottnt{a}  :^{ \ottnt{k} }  \ottnt{A} $.
\end{lemma}

\begin{lemma}[Restriction subsumption]\stack{. }\footnote{\footfilethm{coq/restrict.v}{SubG\_restrict}}
$ \Gamma   \leq    \lceil \Gamma \rceil^{ \ottnt{k} }  $.
\end{lemma}

\subsubsection{Restricted floating}

Subsumption allows variables from one level to be made available
to all higher levels using their current type. 
However, when we use this rule in a judgement, it doesn't
change the context that is used to check the term. This can be
restrictive --- we can only substitute their assumptions with lower level
derivations.

In some cases, we can raise the level
of some assumptions in the context when we raise the level of the judgement
without displacing their types or the rest of the context.
For example, consider the derivable judgement
$   f    :^{ \ottnt{j} }   \Pi \mathit{ \mathit{x} }\!:^{ \ottnt{i} }\! \ottnt{A}  \mathpunct{.}  \ottnt{B}    ,   \mathit{x}    :^{ \ottnt{i} }  \ottnt{A}     \vdash    \mathit{ f }   \;   \mathit{ \mathit{x} }    :^{ \ottnt{j} }  \ottnt{B} $ where $\ottnt{i}  \ottsym{<}  \ottnt{j}$.
We could derive the same judgement at a higher level $ \ottnt{k}  >  \ottnt{j} $
where we also raise the level of $f$ to $\ottnt{k}$.
However, we can only raise the level of variables
at the \emph{same} level as the entire judgement.
In our example, we can't raise $\mathit{x}$ from its lower level $\ottnt{i}$
because then it would be invalid as an argument to $f$.

To prove this formally, we must work with judgements that don't have any
assumptions above the current level by using the restriction operation to discard them.
Next, to raise certain levels, we introduce a \emph{floating} operation 
on contexts $ \mathopen{ \uparrow_{ \ottnt{j} }^{ \ottnt{k} } } \Gamma $
that raises assumptions in $\Gamma$ at level $\ottnt{j}$ to a higher level $\ottnt{k}$
without displacing their types.

\begin{lemma}[Restricted Floating]\stack{. }\footnote{\footfilethm{coq/restrict.v}{DTyping\_float\_restrict}} \label{lem:restricted-floating}
If $ \Delta ;  \Gamma  \vdash  \ottnt{a}  :^{ \ottnt{j} }  \ottnt{A} $ and $\ottnt{j}  \leq  \ottnt{k}$ then $ \Delta ;   \mathopen{ \uparrow_{ \ottnt{j} }^{ \ottnt{k} } } \ottsym{(}   \lceil \Gamma \rceil^{ \ottnt{j} }   \ottsym{)}   \vdash  \ottnt{a}  :^{ \ottnt{k} }  \ottnt{A} $.
\end{lemma}

The restricted floating lemma is required to prove cumulativity of judgements.

\begin{lemma}[Cumulativity]\stack{. }\footnote{\footfilethm{coq/restrict.v}{DTyping\_cumul}} \label{lem:cumulativity}
  If $ \Delta ;  \Gamma  \vdash  \ottnt{a}  :^{ \ottnt{j} }  \ottnt{A} $ and $\ottnt{j}  \leq  \ottnt{k}$ then $ \Delta ;  \Gamma  \vdash  \ottnt{a}  :^{ \ottnt{k} }  \ottnt{A} $.
\end{lemma}

In the nondependent function case $ \Delta ;  \Gamma  \vdash   \lambda \mathit{ \mathit{x} } \mathpunct{.}  \ottnt{b}   :^{ \ottnt{j} }   \ottnt{A}  \mathrel{\rightarrow}  \ottnt{B}  $,
where we want to derive the same judgement at level $ \ottnt{k}  \geq  \ottnt{j} $,
we get by inversion the premise $ \Delta ;   \Gamma ,   \mathit{x}    :^{ \ottnt{j} }  \ottnt{A}     \vdash  \ottnt{b}  :^{ \ottnt{j} }  \ottnt{B} $,
while we need $ \Delta ;   \Gamma ,   \mathit{x}    :^{ \ottnt{k} }  \ottnt{A}     \vdash  \ottnt{b}  :^{ \ottnt{k} }  \ottnt{B} $.
Restricted floating and weakening allows us to raise the level of $\ottnt{b}$ together with the single assumption $\mathit{x}$
from level $\ottnt{j}$ to level $\ottnt{k}$.

\subsubsection{Type Safety}

We can now show that this language satisfies the preservation (\ie subject reduction) and progress lemmas
with respect to call-by-name $\beta\delta$-reduction \fbox{$ \Delta  \vdash  \ottnt{a}  \leadsto^{\ast}  \ottnt{b} $},
whose rules are given in \cref{fig:red}.
For progress, values are type formers and abstractions.

\begin{lemma}[Preservation]\stack{. }\footnote{\footfilethm{coq/typesafety.v}{WHNF\_Preservation}}
\label{lem:preservation}
If $ \Delta ;  \Gamma  \vdash  \ottnt{a}  :^{ \ottnt{k} }  \ottnt{A} $ and $ \Delta  \vdash  \ottnt{a}  \leadsto^{\ast}  \ottnt{a'} $ then $ \Delta ;  \Gamma  \vdash  \ottnt{a'}  :^{ \ottnt{k} }  \ottnt{A} $.
\end{lemma}

\begin{lemma}[Progress]\stack{. }\footnote{\footfilethm{coq/typesafety.v}{WHNF\_Progress}}
If $ \Delta ;  \varnothing  \vdash  \ottnt{a}  :^{ \ottnt{k} }  \ottnt{A} $ then $\ottnt{a}$ is a value or $ \Delta  \vdash  \ottnt{a}  \leadsto^{\ast}  \ottnt{b} $ for some $\ottnt{b}$.
\end{lemma}

\begin{figure}[ht]
  \drule[width=0.5\textwidth]{R-Beta}
  \drule[width=0.5\textwidth]{R-Delta}
  \drule[width=0.5\textwidth]{R-App} \\\\
  \drule[width=0.5\textwidth]{R-Absurd}
  \drule[width=0.5\textwidth]{W-Refl}
  \drule[width=0.5\textwidth]{W-Trans}
\caption{Call-by-name reduction}
\label{fig:red}
\end{figure}

\section{Prototype implementation} \label{sec:implementation}

We have implemented a prototype type checker, which can be found
\ifanonymous in the supplementary materials
\else at \publicrepo{}
\fi
under the \texttt{impl/} directory,
including a brief overview of the concrete syntax\punctstack{.}%
\footnote{\footfile{impl/README.pi}}
This implementation is based on \texttt{pi-forall}~\cite{pi-forall},
a simple bidirectional type checker for a dependently-typed programming language.

\jcxz{These are apparently called ``octagonal'' integer constraints.}
For convenience, displacements and level annotations on dependent types can be omitted;
the type checker then generates level metavariables in their stead.
When checking a single global definition,
constraints on level metavariables are collected,
which form a set of integer inequalities on metavariables.
An SMT solver checks that these inequalities are satisfiable by the naturals
and finally provides a solution that minimizes the levels.
Therefore, assuming the collected constraints are correct,
if a single global definition has a solution,
then a solution will always be found.
However, we don't know if this holds for a \emph{set} of global definitions,
because the solution for a prior definition might affect whether
a later definition that uses it is solveable.
Determining what makes a solution ``better'' or ``more general''
to maximize the number of global definitions that can be solved
is part of future work.

The implementation additionally features stratified datatypes, case expressions, and recursion,
used to demonstrate the practicality of programming in \lang.
Restricting the datatypes to inductive types by checking strict positivity
and termination of recursive functions is possible
but orthogonal to stratification and thus out of scope for this work.
The parameters and arguments of datatypes and their constructors respectively
can be either floating (\ie nondependent) or fixed (\ie dependent),
with their levels following rules analogous to those of nondependent and dependent functions.
Additionally, datatypes and constructors can be displaced like constants,
in that a displaced constructor only belongs to its datatype with the same displacement.

We include with our implementation a small core library\punctstack{,}%
\footnote{\footfile{impl/pi/README.pi}}
and all the examples that appear in this paper have been checked by our implementation\punctstack{.}%
\footnote{\footfile{impl/pi/StraTT.pi}}
In the subsections to follow, we examine three particular datatypes in depth:
decidable types, propositional equality, and dependent pairs.

\subsection{Decidable types}

Revisiting an example from \cref{sec:examples},
we can define $ \mathsf{ Dec } $ as a datatype.
\begin{align*}
  & \mathbf{data} \; \mathsf{ Dec }    {}   \; (\mathit{ X } :   \star  )   :^{ \ottsym{0} }   \star   \; \mathbf{where}  \\
  &\quad  \mathsf{ Yes } :^{ \ottsym{0} }     \mathit{ X }   \mathrel{\rightarrow}   \mathsf{ Dec }    \;   \mathit{ X }    \\
  &\quad  \mathsf{ No } :^{ \ottsym{0} }      \mathsf{ neg }   \;   \mathit{ X }    \mathrel{\rightarrow}   \mathsf{ Dec }    \;   \mathit{ X }   
\end{align*}

The lack of annotation on the parameter indicates that it's a floating domain,
so that $  \lambda \mathit{ X } \mathpunct{.}   \mathsf{ Dec }    \;   \mathit{ X }  $ can be assigned type $  \star   \mathrel{\rightarrow}   \star  $ at level 0.
Datatypes and their constructors, like variables and constants, are cumulative,
so the aforementioned type assignment is valid at any level above 0 as well.
When destructing a datatype, the constructor arguments of each branch are typed such that
the constructor would have the same level as the level of the scrutinee.
Consider the following proof that decidability of a type implies its double negation elimination,
which requires inspecting the decision.
\begin{align*}
  & \mathsf{ decDNE } :^{ \ottsym{1} }       \Pi \mathit{ X }\!:^{ \ottsym{0} }\!  \star   \mathpunct{.}   \mathsf{ Dec }    \;   \mathit{ X }    \mathrel{\rightarrow}   \mathsf{ neg }    \;  \ottsym{(}    \mathsf{ neg }   \;   \mathit{ X }    \ottsym{)}   \mathrel{\rightarrow}   \mathit{ X }    \\
  &     \mathsf{ decDNE }  \;  \mathit{ X }   \;  \mathit{ dec }   \;  \mathit{ nn }    \coloneqq   \mathbf{case} \;   \mathit{ dec }   \; \mathbf{of}   \\
  &\quad    \mathsf{ Yes }  \;  \mathit{ \mathit{y} }    \mathrel{\Rightarrow}   \mathit{ \mathit{y} }   \\
  &\quad    \mathsf{ No }  \;  \mathit{ \mathit{x} }    \mathrel{\Rightarrow}   \mathsf{absurd}(    \mathit{ nn }   \;   \mathit{ \mathit{x} }    )  
\end{align*}

By the level annotation on the function,
we know that $dec$ and $nn$ both have level 1.
Then in the branches, the patterns $  \mathsf{ Yes }   \;   \mathit{ \mathit{y} }  $ and $  \mathsf{ No }   \;   \mathit{ \mathit{x} }  $ must also be typed at level 1,
so that $\mathit{y}$ has type $X$ and $\mathit{x}$ has type $  \mathsf{ neg }   \;   \mathit{ X }  $ both at level 1.

\subsection{Propositional equality}

Datatypes and their constructors, like constants,
can be displaced as well, uniformly raising the levels of their types.
We again revisit an example from \cref{sec:examples}
and now define a propositional equality as a datatype with a single reflexivity constructor.
\begin{align*}
  & \mathbf{data} \; \mathsf{ Eq }    {}   \; (\mathit{ X }   :^{ \ottsym{0} }   \star   )   :^{ \ottsym{1} }     \mathit{ X }   \mathrel{\rightarrow}   \mathit{ X }    \mathrel{\rightarrow}   \star    \; \mathbf{where}  \\
  &\quad  \mathsf{ Refl } :^{ \ottsym{1} }      \Pi \mathit{ \mathit{x} }\!:^{ \ottsym{0} }\!  \mathit{ X }   \mathpunct{.}   \mathit{ Eq }    \;   \mathit{ X }    \;   \mathit{ \mathit{x} }    \;   \mathit{ \mathit{x} }   
\end{align*}

This time, the parameter has a level annotation indicating that it's fixed at $0$,
while its indices are floating.
Displacing $ \mathsf{ Eq } $ by $1$ would then raise the fixed parameter level to $1$,
while the levels of $ \mathsf{ Eq }^{ \ottsym{1} } $ itself and its floating indices always match
but can be $2$ or higher by cumulativity.
Its sole constructor would be $ \mathsf{ Refl }^{ \ottsym{1} } $ containing a single argument of type $X$ at level $1$.
Displacement is needed to state and prove propositions about equalities between equalities,
such as the uniqueness of equality proofs\punctstack{.}%
\footnote{The provability of this principle, also known as UIP~\citep{UIP},
is more a consequence of the quirks of unification in \texttt{pi-forall}
than an intentional intensional design.}
\begin{align*}
  & \mathsf{ UIP } :^{ \ottsym{2} }      \Pi \mathit{ X }\!:^{ \ottsym{0} }\!  \star   \mathpunct{.}   \Pi \mathit{ \mathit{x} }\!:^{ \ottsym{0} }\!  \mathit{ X }   \mathpunct{.}   \Pi \mathit{ p }\!:^{ \ottsym{1} }\!     \mathsf{ Eq }   \;   \mathit{ X }    \;   \mathit{ \mathit{x} }    \;   \mathit{ \mathit{x} }    \mathpunct{.}   \mathsf{ Eq }^{ \ottsym{1} }      \;  \ottsym{(}      \mathsf{ Eq }   \;   \mathit{ X }    \;   \mathit{ \mathit{x} }    \;   \mathit{ \mathit{x} }    \ottsym{)}   \;   \mathit{ p }    \;  \ottsym{(}    \mathsf{ Refl }   \;   \mathit{ \mathit{x} }    \ottsym{)}   \\
  &     \mathsf{ UIP }  \;  \mathit{ X }   \;  \mathit{ \mathit{x} }   \;  \mathit{ p }    \coloneqq    \mathbf{case} \;   \mathit{ p }   \; \mathbf{of}   \;      \mathsf{ Refl }   \;    \mathsf{ \mathit{x} }   \mathrel{\Rightarrow}   \mathsf{ Refl }^{ \ottsym{1} }     \;  \ottsym{(}    \mathsf{ Refl }   \;   \mathit{ \mathit{x} }    \ottsym{)}    
\end{align*}

\subsection{Dependent pairs}

Because there are two different function types,
there are also two different ways to define dependent pairs.
Using a floating function type for the second component's type
results in pairs whose first and second projections can be defined as usual,
while using the stratified dependent function type results in pairs
whose second projection can't be defined in terms of the first.
We first take a look at the former.
\begin{align*}
  & \mathbf{data} \; \mathsf{ NPair }     {}   \; (\mathit{ X }   :^{ \ottsym{0} }   \star   )   \; (\mathit{ P } :    \mathit{ X }   \mathrel{\rightarrow}   \star   )   :^{ \ottsym{1} }   \star   \; \mathbf{where}  \\
  &\quad  \mathsf{ MkPair } :^{ \ottsym{1} }       \Pi \mathit{ \mathit{x} }\!:^{ \ottsym{0} }\!  \mathit{ X }   \mathpunct{.}   \mathit{ P }    \;   \mathit{ \mathit{x} }    \mathrel{\rightarrow}   \mathsf{ NPair }    \;   \mathit{ X }    \;   \mathit{ P }    \\
  & \mathsf{ nfst } :^{ \ottsym{1} }      \Pi \mathit{ X }\!:^{ \ottsym{0} }\!  \star   \mathpunct{.}   \Pi \mathit{ P }\!:^{ \ottsym{0} }\!   \mathit{ X }   \mathrel{\rightarrow}   \star    \mathpunct{.}   \mathsf{ NPair }     \;   \mathit{ X }    \;   \mathit{ P }    \mathrel{\rightarrow}   \mathit{ X }    \\
  &     \mathsf{ nfst }  \;  \mathit{ X }   \;  \mathit{ P }   \;  \mathit{ p }    \coloneqq   \mathbf{case} \;   \mathit{ p }   \; \mathbf{of}   \;     \mathsf{ MkPair }  \;  \mathit{ \mathit{x} }   \;  \mathit{ \mathit{y} }    \mathrel{\Rightarrow}   \mathit{ \mathit{x} }   \\
  & \mathsf{ nsnd } :^{ \ottsym{2} }    \Pi \mathit{ X }\!:^{ \ottsym{0} }\!  \star   \mathpunct{.}   \Pi \mathit{ P }\!:^{ \ottsym{0} }\!   \mathit{ X }   \mathrel{\rightarrow}   \star    \mathpunct{.}   \Pi \mathit{ p }\!:^{ \ottsym{1} }\!    \mathsf{ NPair }   \;   \mathit{ X }    \;   \mathit{ P }    \mathpunct{.}   \mathit{ P }      \;  \ottsym{(}      \mathsf{ nfst }   \;   \mathit{ X }    \;   \mathit{ P }    \;   \mathit{ p }    \ottsym{)}   \\
  &     \mathsf{ nsnd }  \;  \mathit{ X }   \;  \mathit{ P }   \;  \mathit{ p }    \coloneqq   \mathbf{case} \;   \mathit{ p }   \; \mathbf{of}   \;     \mathsf{ MkPair }  \;  \mathit{ \mathit{x} }   \;  \mathit{ \mathit{y} }    \mathrel{\Rightarrow}   \mathit{ \mathit{y} }  
\end{align*}

Due to stratification, the projections need to be defined at level 1 and 2 respectively
to accommodate dependently quantifying over the parameters at level $0$
and the pair at level $1$.
Even so, the second projection is well typed,
since $P$ can be used at level $2$ by subsumption
to be applied to the first projection at level $2$ also by subsumption
in the return type of the second projection.

As the two function types are distinct,
we do need both varieties of dependent pairs.
In particular, with the above pairs alone, we aren't able to type check a universe of propositions
$   \mathsf{ NPair }   \;   \star    \;   \textcolor{red}{ \underline{  \mathsf{ isProp }  } }  $, as the predicate has type $ \Pi \mathit{ X }\!:^{ \ottsym{0} }\!  \star   \mathpunct{.}   \star  $ at level 1.
\begin{align*}
  & \mathbf{data} \; \mathsf{ DPair }     {}   \; (\mathit{ X }   :^{ \ottsym{0} }   \star   )   \; (\mathit{ P } :   \Pi \mathit{ \mathit{x} }\!:^{ \ottsym{0} }\!  \mathit{ X }   \mathpunct{.}   \star   )   :^{ \ottsym{1} }   \star   \; \mathbf{where}  \\
  &\quad  \mathsf{ MkPair } :^{ \ottsym{1} }       \Pi \mathit{ \mathit{x} }\!:^{ \ottsym{0} }\!  \mathit{ X }   \mathpunct{.}   \mathit{ P }    \;   \mathit{ \mathit{x} }    \mathrel{\rightarrow}   \mathsf{ DPair }    \;   \mathit{ X }    \;   \mathit{ P }    \\
  & \mathsf{ dfst } :^{ \ottsym{2} }      \Pi \mathit{ X }\!:^{ \ottsym{0} }\!  \star   \mathpunct{.}   \Pi \mathit{ P }\!:^{ \ottsym{1} }\! \ottsym{(}   \Pi \mathit{ \mathit{x} }\!:^{ \ottsym{0} }\!  \mathit{ X }   \mathpunct{.}   \star    \ottsym{)}  \mathpunct{.}   \mathsf{ DPair }     \;   \mathit{ X }    \;   \mathit{ P }    \mathrel{\rightarrow}   \mathit{ X }    \\
  &     \mathsf{ dfst }  \;  \mathit{ X }   \;  \mathit{ P }   \;  \mathit{ p }    \coloneqq   \mathbf{case} \;   \mathit{ p }   \; \mathbf{of}   \;     \mathsf{ MkPair }  \;  \mathit{ \mathit{x} }   \;  \mathit{ \mathit{y} }    \mathrel{\Rightarrow}   \mathit{ \mathit{x} }   \\
  & \mathsf{ dsnd } :^{ \ottsym{2} }   \Pi \mathit{ X }\!:^{ \ottsym{0} }\!  \star   \mathpunct{.}   \Pi \mathit{ P }\!:^{ \ottsym{1} }\! \ottsym{(}   \Pi \mathit{ \mathit{x} }\!:^{ \ottsym{0} }\!  \mathit{ X }   \mathpunct{.}   \star    \ottsym{)}  \mathpunct{.}   \Pi \mathit{ p }\!:^{ \ottsym{1} }\!    \mathsf{ DPair }   \;   \mathit{ X }    \;   \mathit{ P }    \mathpunct{.}   {}      \\
  &\phantom{dsnd \mathrel{:^{2}}}  \mathbf{case} \;   \mathit{ p }   \; \mathbf{of}  \;      \mathsf{ MkPair }  \;  \mathit{ \mathit{x} }   \;  \mathit{ \mathit{y} }    \mathrel{\Rightarrow}   \mathit{ P }    \;   \mathit{ \mathit{x} }   \\
  &     \mathsf{ dsnd }  \;  \mathit{ X }   \;  \mathit{ P }   \;  \mathit{ p }    \coloneqq   \mathbf{case} \;   \mathit{ p }   \; \mathbf{of}   \;     \mathsf{ MkPair }  \;  \mathit{ \mathit{x} }   \;  \mathit{ \mathit{y} }    \mathrel{\Rightarrow}   \mathit{ \mathit{y} }  
\end{align*}

In the second variant of dependent pairs where $P$ is a stratified dependent function type,
the domain of $P$ is fixed to level 0,
so in the type in $ \mathsf{ dsnd } $, it can't be applied to the first projection,
but it can still be applied to the first component by matching on the pair.
Now we're able to type check $   \mathsf{ DPair }   \;   \star    \;   \mathsf{ isProp }  $.

In both cases, the first component has a fixed level,
while the second component is floating,
so using a predicate at a higher level results in a pair type at a higher level by subsumption.
Consider the predicate $ \mathsf{ isSet } $, which has type $ \Pi \mathit{ X }\!:^{ \ottsym{0} }\!  \star   \mathpunct{.}   \star  $ at level 2:
the universe of sets $   \mathsf{ DPair }   \;   \star    \;   \mathsf{ isSet }  $ is also well typed at level 2.

Unfortunately, the first projection $ \mathsf{ dfst } $ can no longer be used on an element of this pair,
since the predicate is now at level 2,
nor can its displacement $ \mathsf{ dfst }^{ \ottsym{1} } $, since that would displace the level of the first component as well.
Without proper level polymorphism, which would allow keeping the first argument's level fixed
while setting the second argument's level to 2,
we're forced to write a whole new first projection function.

In general, this limitation occurs whenever a datatype contains both dependent and nondependent parameters.
Nevertheless, in the case of the pair type,
the flexibility of a nondependent second component type is still preferable
to a dependent one that fixes its level,
since there would need to be entirely separate datatype definitions
for different combinations of first and second component levels,
\ie one with levels 0 and 1 (as in the case of $ \mathsf{ isProp } $),
one with levels 0 and 2 (as in the case of $ \mathsf{ isSet } $),
and so on.

\section{Discussion} \label{sec:discussion}

\subsection{On consistency}

The consistency of \sublang tells us that the basic premise
of using stratification in place of a universe hierarchy is sensible.
However, it isn't necessarily an incremental step towards consistency of the full \lang,
as we've seen that directly adding floating functions to the logical relation doesn't work,
and an entirely different approach may be needed after all.

One possible direction is to take inspiration from the syntactic metatheory,
especially the \nameref{lem:restricted-floating} lemma, which is required specifically to show cumulativity of floating functions.
Since cumulativity is exactly where the na\"ive addition of floating functions to the logical relation fails,
the key may be to formulate this lemma semantically.
This might require modifying the logical relation to involve contexts and to relate open terms instead.

Another possibility is based on the observation that due to cumulativity,
floating functions appear to be parametric in its stratification level,
at least starting from the smallest level at which it can be well typed.
This suggests that some sort of relational model may help to interpret levels parametrically.

Nevertheless, we strongly believe that \lang is indeed consistent.
The \nameref{lem:restriction} lemma in particular intuitively tells us
that nothing at higher levels could possibly be smuggled into a lower level to violate stratification.
As a further confidence check, we have verified that three type-theoretic paradoxes
possible in an ordinary type theory with type-in-type
do \emph{not} type check in our implementation.
These paradoxes are Burali-Forti's paradox~\citep{burali-forti},
Russell's paradox~\citep{russell},
and Hurkens' paradox~\citep{hurkens},
which all end up reaching a point where a higher-level term
needs to fit into a lower-level position to proceed any further
--- exactly what stratification is designed to prevent.
\Cref{sec:appx:paradoxes} examines these paradoxes in depth.

\subsection{On useability}

Useability comes down to the balance between practicality and expressivity.
On the practicality side, our implementation demonstrates that if a definition is well typed,
then its levels and displacements can be completely omitted and inferred,
a workflow comparable to Coq or Lean.
Additionally, since constants are displaced by only a single displacement,
\lang doesn't exhibit the same kind of exponential blowup in levels and type checking time
that can occur when using universe-polymorphic definitions in Coq or Lean,
which need to abstract over and instantiate over all implicit levels involved.
This behaviour is demonstrated by the concrete, though artificial, examples in \cref{sec:appx:upoly},
whose corresponding \lang definition checks just fine\punctstack{.}%
\footnote{\footfile{impl/pi/Blowup.pi}}
However, if a definition is \emph{not} well typed,
debugging it may involve wading through constraints between generated level metavariables
in situations normally having nothing to do with universe levels,
since stratification now involves levels everywhere,
in particular when using dependent function types.

On the expressivity side,
the displacement system of \lang falls somewhere between level monomorphism
and prenex level polymorphism;
in some scenarios, it works just as well as polymorphism.
For instance, to type check Hurkens' paradox as far as \lang can,
the Coq formulation of the paradox without type-in-type
requires turning on universe polymorphism,
and the Agda formulation of the paradox without type-in-type
requires definitions polymorphic over at least three universe levels.
In general, displacement seems particularly suited for our stratified system,
since level annotations only appear on dependent function domains, not on universes.
For example, the type $  \Pi \mathit{ X }\!:^{ \ottsym{0} }\!  \star   \mathpunct{.}  \ottsym{(}    \mathit{ X }   \mathrel{\rightarrow}   \star    \ottsym{)}   \mathrel{\rightarrow}   \star  $ only has one level,
while the corresponding most general Agda type
\mintinline{agda}|(X : Set ℓ₁) → (X → Set ℓ₂) → Set ℓ₃|
has three and would fare poorly with displacement.

However, in other scenarios, the expressivity of level polymorphism
over multiple level variables is truly needed.
For instance, merely having a type constructor with both a dependent domain
and a nondependent domain interacts poorly with cumulativity.
Suppose we had some type constructor $ \mathsf{ T } :^{ \ottsym{1} }    \Pi \mathit{ \mathit{x} }\!:^{ \ottsym{0} }\!  \mathit{ X }   \mathpunct{.}   \mathit{ Y }    \mathrel{\rightarrow}   \star   $
and a function over elements of this type $ \mathsf{ f } :^{ \ottsym{1} }      \Pi \mathit{ \mathit{x} }\!:^{ \ottsym{0} }\!  \mathit{ X }   \mathpunct{.}   \Pi \mathit{ \mathit{y} }\!:^{ \ottsym{0} }\!  \mathit{ Y }   \mathpunct{.}   \mathsf{ T }     \;   \mathit{ \mathit{x} }    \;   \mathit{ \mathit{y} }    \mathrel{\rightarrow}   \mathit{ Z }   $.
By cumulativity, if $\mathit{y}$ has level $\ottsym{2}$, $   \mathsf{ T }   \;   \mathit{ \mathit{x} }    \;   \mathit{ \mathit{y} }  $ is still well typed by cumulativity at level $\ottsym{2}$,
but $ \mathsf{ f } $ can no longer be applied to it, since the level of $\mathit{y}$ is now too high.
We would like the second argument of $ \mathsf{ f } $ to float along with $ \mathsf{ T } $,
but this isn't possible since it's depended upon.
Having the level of the second argument be polymorphic (subject to the expected constraints)
would resolve this issue.

\subsection{Related work}

\lang is directly inspired from Leivant's stratified polymorphism~\citep{stratified-systemf, finitely-stratified-systemf, transfinite-stratified-systemf},
which developed from Statman's ramified polymorphic typed $\lambda$-calculus~\citep{RPTL}.
Stratified System F, a slight modification of the original system,
has since been used to demonstrate a normalization proof technique using hereditary substitution~\citep{hereditary-systemf},
which in turn has been mechanized in Coq as a case study for the Equations package~\citep{equations-hereditary-systemf}.
More recently, an interpreter of an intrinsically-typed Stratified System F
has been mechanized in Agda by Thiemann and Weidner~\citep{interp-stratified-systemf},
where stratification levels are interpreted as Agda's universe levels.
Similarly, Hubers and Morris' Stratified $\mathrm{R}_\omega$,
a stratified System $\mathrm{F}_\omega$ with row types,
has been mechanized in Agda as well~\citep{stratified-row-omega}.
Meanwhile, our system of level displacement comes from McBride's \crude stratification~\citep{crude, crude-slides},
specializing the displacement algebra (in the sense of Favonia, Angiuli, and Mullanix~\citep{displacement}) to the naturals.

\section{Conclusion} \label{sec:conclusion}

In this work, we have introduced \langue,
a departure from a decades-old tradition of universe hierarchies without,
we believe, succumbing to the threat of logical inconsistency.
By stratifying dependent function types,
we obstruct the usual avenues by which paradoxes manifest their inconsistencies;
and by separately introducing floating nondependent function types,
we recover some of the expressivity lost under the strict rule of stratification.
Although proving logical consistency for the full \lang remains future work,
we \emph{have} proven it for the subsystem \sublang,
and we have provided supporting evidence
by showing how well-known type-theoretic paradoxes fail.

Towards demonstrating that \lang isn't a mere theoretical exercise
and, if consistent, is a viable basis for theorem proving and dependently-typed programming,
we have implemented a prototype type checker for the language
augmented with datatypes, along with a small core library.
The implementation also features inference for level annotations and displacements,
allowing the user to omit them entirely.
We leave formally ensuring that our rules for datatypes
don't violate existing metatheoretical properties as future work as well.

Given the various useability tradeoffs discussed,
as well as the incomplete status of its consistency,
we don't see any particularly compelling reason for existing proof assistants
to adopt a system based on \lang,
but we don't anticipate any particular showstoppers, either,
and believe it suitable for further improvement and iteration.
Ultimately, we hope that \lang demonstrates the feasibility of a renewed alternative to how type universes are handled,
and opens up fresh avenues in the design space of type theories for proof assistants.

\bibliographystyle{plainurl}
\bibliography{refs}

\appendix

\section{Paradoxes} \label{sec:appx:paradoxes}

\subsection{Burali-Forti's paradox}

Burali-Forti's paradox~\citep{burali-forti} in set theory concerns
the simultaneous \wellfoundedness and non--\wellfoundedness of an ordinal.
In type theory,
we instead consider a particular datatype $ \mathsf{ U } $ due to Coquand~\citep{trees}\punctstack{,}%
\footnote{Our thanks to Stephen Dolan for detailing to us this example.}%
\textsuperscript{,}%
\footnote{\footfile{impl/pi/WFU.pi}}
along with a \wellfoundedness predicate for $ \mathsf{ U } $.
\begin{align*}
  & \mathbf{data} \; \mathsf{ U }   {}   :^{ \ottsym{1} }   \star   \; \mathbf{where}  \\
  &\quad  \mathsf{ MkU } :^{ \ottsym{1} }    \Pi \mathit{ X }\!:^{ \ottsym{0} }\!  \star   \mathpunct{.}  \ottsym{(}    \mathit{ X }   \mathrel{\rightarrow}   \mathsf{ U }    \ottsym{)}   \mathrel{\rightarrow}   \mathsf{ U }    \\
  & \mathbf{data} \; \mathsf{ WF }   {}   :^{ \ottsym{2} }    \mathsf{ U }   \mathrel{\rightarrow}   \star    \; \mathbf{where}  \\
  &\quad  \mathsf{ MkWF } :^{ \ottsym{2} }     \Pi \mathit{ X }\!:^{ \ottsym{0} }\!  \star   \mathpunct{.}   \Pi \mathit{ f }\!:^{ \ottsym{1} }\!   \mathit{ X }   \mathrel{\rightarrow}   \mathsf{ U }    \mathpunct{.}  \ottsym{(}    \Pi \mathit{ \mathit{x} }\!:^{ \ottsym{1} }\!  \mathit{ X }   \mathpunct{.}   \mathsf{ WF }    \;  \ottsym{(}    \mathit{ f }   \;   \mathit{ \mathit{x} }    \ottsym{)}   \ottsym{)}    \mathrel{\rightarrow}   \mathsf{ WF }    \;  \ottsym{(}     \mathsf{ MkU }   \;   \mathit{ X }    \;   \mathit{ f }    \ottsym{)}  
\end{align*}

Note that both of these definitions are strictly positive,
so we aren't using any tricks relying on negative datatypes.
It's easy to show that all $ \mathsf{ U } $ are well founded.
\begin{align*}
  & \mathsf{ wf } :^{ \ottsym{2} }    \Pi \mathit{ u }\!:^{ \ottsym{1} }\!   \mathsf{ U }    \mathpunct{.}   \mathsf{ WF }    \;   \mathit{ u }    \\
  &   \mathsf{ wf }  \;  \mathit{ u }    \coloneqq   \mathbf{case} \;   \mathit{ u }   \; \mathbf{of}   \\
  &\quad        \mathsf{ MkU }  \;  \mathit{ X }   \;  \mathit{ f }    \mathrel{\Rightarrow}   \mathsf{ MkWF }    \;   \mathit{ X }    \;   \mathit{ f }    \;  \ottsym{(}    \lambda \mathit{ \mathit{x} } \mathpunct{.}   \mathsf{ wf }    \;  \ottsym{(}    \mathit{ f }   \;   \mathit{ \mathit{x} }    \ottsym{)}   \ottsym{)} 
\end{align*}

The usual paradox, with type-in-type and without stratification,
constructs a $ \mathsf{ U } $ that is provably \emph{not} well founded.
\begin{align*}
  & \mathsf{ loop } :^{ \ottsym{1} }   \mathsf{ U }   \\
  &  \mathsf{ loop }   \coloneqq     \mathsf{ MkU }   \;   \textcolor{red}{ \underline{  \mathsf{ U }  } }    \;  \ottsym{(}   \lambda \mathit{ u } \mathpunct{.}   \mathit{ u }    \ottsym{)}   \\
  & \mathsf{ nwfLoop } :^{ \ottsym{2} }     \mathsf{ WF }   \;   \mathsf{ loop }    \mathrel{\rightarrow}   \bot    \\
  &   \mathsf{ nwfLoop }  \;  \mathit{ wfLoop }    \coloneqq   \mathbf{case} \;   \mathit{ wfLoop }   \; \mathbf{of}   \\
  &\quad       \mathsf{ MkWF }  \;  \mathit{ X }   \;  \mathit{ f }   \;  \mathit{ h }    \mathrel{\Rightarrow}   \mathsf{ nwfLoop }    \;  \ottsym{(}    \mathit{ h }   \;   \mathsf{ loop }    \ottsym{)} 
\end{align*}

In the branch of $ \mathsf{ nwfLoop } $, by pattern matching on the type of the scrutinee,
$X$ is bound to $ \mathsf{ U } $ and $f$ to $ \lambda \mathit{ u } \mathpunct{.}   \mathit{ u }  $,
so $  \mathit{ h }   \;   \mathsf{ loop }  $ correctly has type $  \mathsf{ WF }   \;    \mathsf{ loop }   $.
Note that this definition would also pass the usual structural termination check,
since the recursive call is done on a subargument from $h$.
Then $  \mathsf{ nwfLoop }   \;  \ottsym{(}    \mathsf{ wf }   \;    \mathsf{ loop }     \ottsym{)} $ is an inhabitant of the empty type.

With stratification, $ \mathsf{ U } $ with level $1$ is too large
to fit into the type argument of $ \mathsf{ MkU } $, which demands level $0$,
so $ \mathsf{ loop } $ can't be constructed in the first place.
This is also why the level of a datatype can't be strictly lower than
that of its constructors, despite such a design not violating the regularity lemma for constructors.

\subsection{Russell's paradox}

The $ \mathsf{ U } $ above was originally used by Coquand~\citep{trees}
to express a variant of Russell's paradox~\citep{russell}\punctstack{.}%
\footnote{An Agda implementation can be found at \\ \url{https://github.com/agda/agda/blob/master/test/Succeed/Russell.agda}~\citep{capriotti}.}%
\textsuperscript{,}%
\footnote{\footfile{impl/pi/Russell.pi}}
First, a $ \mathsf{ U } $ is said to be regular if it's provably inequal to its subarguments;
this represents a set which doesn't contain itself.
\begin{align*}
  & \mathsf{ regular } :^{ \ottsym{1} }    \mathsf{ U }   \mathrel{\rightarrow}   \star    \\
  &   \mathsf{ regular }  \;  \mathit{ u }    \coloneqq   \mathbf{case} \;   \mathit{ u }   \; \mathbf{of}   \\
  &\quad      \mathsf{ MkU }  \;  \mathit{ X }   \;  \mathit{ f }    \mathrel{\Rightarrow}   \Pi \mathit{ \mathit{x} }\!:^{ \ottsym{0} }\!  \mathit{ X }   \mathpunct{.}  \ottsym{(}       \mathit{ f }   \;   \mathit{ \mathit{x} }    =   \mathsf{ MkU }    \;   \mathit{ X }    \;   \mathit{ f }    \ottsym{)}    \mathrel{\rightarrow}   \bot  
\end{align*}

The trick is to define a $ \mathsf{ U } $ that is both regular and nonregular.
Normally, with type-in-type, this would be one that represents
the set of all regular sets.
\begin{align*}
  & \mathsf{ R } :^{ \ottsym{3} }   \mathsf{ U }^{ \ottsym{2} }   \\
  &  \mathsf{ R }   \coloneqq     \mathsf{ MkU }^{ \ottsym{2} }   \;  \ottsym{(}     \mathsf{ NPair }^{ \ottsym{1} }   \;   \mathsf{ U }    \;   \mathsf{ regular }    \ottsym{)}   \;   \textcolor{red}{ \underline{ \ottsym{(}     \mathsf{ nfst }^{ \ottsym{1} }   \;   \mathsf{ U }    \;   \mathsf{ regular }    \ottsym{)} } }   
\end{align*}

Stratification once again prevents $ \mathsf{ R } $ from type checking,
since the pair projection returns a $ \mathsf{ U } $ and not a $ \mathsf{ U }^{ \ottsym{2} } $.
The type contained in the pair can't be displaced to $ \mathsf{ U }^{ \ottsym{2} } $ either,
since that would make the pair's level too large to fit inside $ \mathsf{ MkU }^{ \ottsym{2} } $\stackpunct{2}{.}

\subsection{Hurkens' paradox}

Although we've seen that stratification thwarts the paradoxes above,
they leverage the properties of datatypes and recursive functions,
which we haven't formalized.
Here, we'll turn to the failure of Hurkens' paradox~\citep{hurkens}
as further evidence of consistency,
which in contrast can be formulated in pure \lang without datatypes.
Below is the paradox in Coq without universe checking.

\begin{minted}{coq}
  Require Import Coq.Unicode.Utf8_core.
  Unset Universe Checking.
  Definition P (X : Type) : Type := X → Type.
  Definition U : Type :=
    ∀ (X : Type), (P (P X) → X) → P (P X).
  Definition tau (t : P (P U)) : U :=
    λ X f p, t (λ s, p (f (s X f))).
  Definition sig (s : U) : P (P U) := s U tau.
  Definition Delta (y : U) : Type :=
    (∀ (p : P U), sig y p → p (tau (sig y))) → False.
  Definition Omega : U :=
    tau (λ p, ∀ (x : U), sig x p → p x).
  Definition M (x : U) (s : sig x Delta) : Delta x :=
    λ d, d Delta s (λ p, d (λ y, p (tau (sig y)))).
  Definition D := ∀ p, (∀ x, sig x p → p x) → p Omega.
  Definition R : D :=
    λ p d, d Omega (λ y, d (tau (sig y))).
  Definition L (d : D) : False :=
    d Delta M (λ p, d (λ y, p (tau (sig y)))).
  Definition false : False := L R.
\end{minted}

If we replace unsetting universe checking with
\begin{minted}{coq}
  Set Universe Polymorphism.
\end{minted}
then the definitions check up to \mintinline{coq}{M}.
Similarly, in Agda, we can get the paradox to check up to \mintinline{agda}{M}
by using explicit universe polymorphism.

\begin{minted}{agda}
  {-# OPTIONS --cumulativity #-}
  open import Agda.Primitive

  data ⊥ : Set where

  U : ∀ ℓ ℓ₁ ℓ₂ → Set (lsuc (ℓ ⊔ ℓ₁ ⊔ ℓ₂))
  U ℓ ℓ₁ ℓ₂ = ∀ (X : Set ℓ) → (((X → Set ℓ₁) → Set ℓ₂) → X) → ((X → Set ℓ₁) → Set ℓ₂) 

  τ : ∀ ℓ₁ ℓ₂ → ((U ℓ₁ ℓ₁ ℓ₂ → Set ℓ₁) → Set ℓ₂) → U ℓ₁ ℓ₁ ℓ₂
  τ ℓ₁ ℓ₂ t = λ X f p → t (λ x → p (f (x X f)))

  σ : ∀ ℓ₁ ℓ₂ → U (lsuc (ℓ₁ ⊔ ℓ₂)) ℓ₁ ℓ₂ → (U ℓ₁ ℓ₁ ℓ₂ → Set ℓ₁) → Set ℓ₂
  σ ℓ₁ ℓ₂ s = s (U ℓ₁ ℓ₁ ℓ₂) (τ ℓ₁ ℓ₂)

  Δ : ∀ {ℓ₁ ℓ₂} → U (lsuc (ℓ₁ ⊔ ℓ₂)) ℓ₁ ℓ₂ → Set (lsuc (ℓ₁ ⊔ ℓ₂))
  Δ {ℓ₁} {ℓ₂} y = (∀ p → σ ℓ₁ ℓ₂ y p → p (τ ℓ₁ ℓ₂ (σ ℓ₁ ℓ₂ y))) → ⊥

  Ω : ∀ {ℓ} → U ℓ ℓ (lsuc (lsuc ℓ))
  Ω {ℓ} = τ ℓ (lsuc (lsuc ℓ)) (λ p → (∀ x → σ ℓ ℓ x p → p x))

  M : ∀ {ℓ} x → σ (lsuc ℓ) ℓ x (Δ {ℓ} {ℓ}) → Δ {lsuc ℓ} {ℓ} x
  M {ℓ} _ 𝟚 𝟛 = 𝟛 Δ 𝟚 (λ p → 𝟛 (λ y → p (τ ℓ ℓ (σ ℓ ℓ y))))

  R : ∀ {ℓ} p → (∀ x → σ ℓ (lsuc (lsuc ℓ)) x p → p x) → p Ω
  R {ℓ} _ 𝟙 = {! 𝟙 (Ω {ℓ}) (λ x → 𝟙 (τ ℓ ℓ (σ ℓ ℓ x))) !}
  -- Need Ω : U (lsuc (lsuc (lsuc ℓ))) ℓ (lsuc (lsuc ℓ))
  -- Have Ω : U ℓ ℓ (lsuc (lsuc ℓ))

  L : ∀ {ℓ} → (∀ p → (∀ x → σ ℓ (lsuc (lsuc ℓ)) x p → p x) → p Ω) → ⊥
  L {ℓ} 𝟘 = {! 𝟘 (Δ {ℓ} {ℓ}) M (λ p → 𝟘 (λ y → p (τ ℓ ℓ ℓ (σ ℓ ℓ ℓ y)))) !}
  -- Need Δ : U ℓ ℓ (lsuc (lsuc ℓ)) → Set ℓ
  -- Have Δ : U (lsuc ℓ) ℓ ℓ → Set (lsuc ℓ)

  false : ⊥
  false = L {lzero} (R {lzero})
\end{minted}

The corresponding \lang code, too, checks up to $ \mathsf{ M } $,
as verified in the implementation\punctstack{.}%
\footnote{\footfile{impl/pi/Hurkens.pi} (no annotations),
\footfile{impl/pi/HurkensAnnot.pi} (all annotations)}
Displacement is sufficient to handle situations in which polymorphism was needed.
\begin{align*}
  & \mathsf{ P } :^{ \ottsym{0} }    \star   \mathrel{\rightarrow}   \star    \\
  &   \mathsf{ P }  \;  \mathit{ X }    \coloneqq    \mathit{ X }   \mathrel{\rightarrow}   \star    \\
  & \mathsf{ U } :^{ \ottsym{1} }   \star   \\
  &  \mathsf{ U }   \coloneqq     \Pi \mathit{ X }\!:^{ \ottsym{0} }\!  \star   \mathpunct{.}  \ottsym{(}     \mathsf{ P }   \;  \ottsym{(}    \mathsf{ P }   \;   \mathit{ X }    \ottsym{)}   \mathrel{\rightarrow}   \mathit{ X }    \ottsym{)}   \mathrel{\rightarrow}   \mathsf{ P }    \;  \ottsym{(}    \mathsf{ P }   \;   \mathit{ X }    \ottsym{)}   \\
  & \mathsf{ tau } :^{ \ottsym{1} }     \mathsf{ P }   \;  \ottsym{(}    \mathsf{ P }   \;   \mathsf{ U }    \ottsym{)}   \mathrel{\rightarrow}   \mathsf{ U }    \\
  &      \mathsf{ tau }  \;  \mathit{ t }   \;  \mathit{ X }   \;  \mathit{ f }   \;  \mathit{ p }    \coloneqq    \mathit{ t }   \;  \ottsym{(}    \lambda \mathit{ s } \mathpunct{.}   \mathit{ p }    \;  \ottsym{(}    \mathit{ f }   \;  \ottsym{(}     \mathit{ s }   \;   \mathit{ X }    \;   \mathit{ f }    \ottsym{)}   \ottsym{)}   \ottsym{)}   \\
  & \mathsf{ sig } :^{ \ottsym{2} }     \mathsf{ U }^{ \ottsym{1} }   \mathrel{\rightarrow}   \mathsf{ P }    \;  \ottsym{(}    \mathsf{ P }   \;   \mathsf{ U }    \ottsym{)}   \\
  &   \mathsf{ sig }  \;  \mathit{ s }    \coloneqq     \mathit{ s }   \;    \mathsf{ U }     \;    \mathsf{ tau }     \\
  & \mathsf{ Delta } :^{ \ottsym{2} }    \mathsf{ P }   \;   \mathsf{ U }^{ \ottsym{1} }    \\
  &   \mathsf{ Delta }  \;  \mathit{ \mathit{y} }    \coloneqq   \ottsym{(}       \Pi \mathit{ p }\!:^{ \ottsym{1} }\!   \mathsf{ P }   \;   \mathsf{ U }    \mathpunct{.}   \mathsf{ sig }    \;   \mathit{ \mathit{y} }    \;   \mathit{ p }    \mathrel{\rightarrow}   \mathit{ p }    \;  \ottsym{(}    \mathsf{ tau }   \;  \ottsym{(}    \mathsf{ sig }   \;   \mathit{ \mathit{y} }    \ottsym{)}   \ottsym{)}   \ottsym{)}  \mathrel{\rightarrow}   \bot    \\
  & \mathsf{ Omega } :^{ \ottsym{3} }   \mathsf{ U }   \\
  &  \mathsf{ Omega }   \coloneqq    \mathsf{ tau }   \;  \ottsym{(}       \lambda \mathit{ p } \mathpunct{.}   \Pi \mathit{ \mathit{x} }\!:^{ \ottsym{2} }\!  \mathsf{ U }^{ \ottsym{1} }   \mathpunct{.}   \mathsf{ sig }     \;   \mathit{ \mathit{x} }    \;   \mathit{ p }    \mathrel{\rightarrow}   \mathit{ p }    \;  \ottsym{(}    \lambda \mathit{ X } \mathpunct{.}   \mathit{ \mathit{x} }    \;   \mathit{ X }    \ottsym{)}   \ottsym{)}   \\
  & \mathsf{ M } :^{ \ottsym{4} }       \Pi \mathit{ \mathit{x} }\!:^{ \ottsym{3} }\!  \mathsf{ U }^{ \ottsym{2} }   \mathpunct{.}   \mathsf{ sig }^{ \ottsym{1} }    \;   \mathit{ \mathit{x} }    \;   \mathsf{ Delta }    \mathrel{\rightarrow}   \mathsf{ Delta }^{ \ottsym{1} }    \;   \mathit{ \mathit{x} }    \\
  &     \mathsf{ M }  \;  \mathit{ \mathit{x} }   \;  \mathit{ s }   \;  \mathit{ d }    \coloneqq      \mathit{ d }   \;   \mathsf{ Delta }    \;   \mathit{ s }    \;  \ottsym{(}    \lambda \mathit{ p } \mathpunct{.}   \mathit{ d }    \;  \ottsym{(}    \lambda \mathit{ \mathit{y} } \mathpunct{.}   \mathit{ p }    \;  \ottsym{(}    \mathsf{ tau }   \;  \ottsym{(}    \mathsf{ sig }   \;   \mathit{ \mathit{y} }    \ottsym{)}   \ottsym{)}   \ottsym{)}   \ottsym{)}   \\
  & \mathsf{ D } :^{ \ottsym{3} }   \star   \\
  &  \mathsf{ D }   \coloneqq     \Pi \mathit{ p }\!:^{ \ottsym{1} }\!   \mathsf{ P }   \;   \mathsf{ U }    \mathpunct{.}  \ottsym{(}       \Pi \mathit{ \mathit{x} }\!:^{ \ottsym{1} }\!  \mathsf{ U }   \mathpunct{.}   \mathsf{ sig }    \;   \textcolor{red}{ \underline{  \mathit{ \mathit{x} }  } }    \;   \mathit{ p }    \mathrel{\rightarrow}   \mathit{ p }    \;   \mathit{ \mathit{x} }    \ottsym{)}   \mathrel{\rightarrow}   \mathit{ p }    \;   \mathsf{ Omega }   
\end{align*}

The next definition $ \mathsf{ D } $ doesn't type check,
since $ \mathsf{ sig } $ takes a displaced $ \mathsf{ U }^{ \ottsym{1} } $ and not a $ \mathsf{ U } $.
The type of $\mathit{x}$ can't be displaced to fix this either,
since $p$ takes an undisplaced $ \mathsf{ U } $ and not a $ \mathsf{ U }^{ \ottsym{1} } $.
Being stuck trying to equate two different levels is reassuring,
as conflating different universe levels is how we expect
a paradox that exploits type-in-type to operate.

\subsection{Reynolds' paradox}

Our final example concerns the inconsistency of inductives
which are positive but not \emph{strictly} so
together with an impredicative universe,
as described by Coquand and Paulin-Mohring~\citep{inductives}\punctstack{.}%
\footnote{A Coq implementation has been made by Sj\"oberg~\citep{spit}.}%
\textsuperscript{,}%
\footnote{\footfile{impl/pi/Reynolds.pi}}
We consider such a nonstrictly-positive datatype $ \mathsf{A_0} $.
\begin{align*}
  & \mathbf{data} \; \mathsf{A_0}   {}   :^{ \ottsym{0} }   \star   \; \mathbf{where}  \\
  &\quad  \mathsf{A_0} :^{ \ottsym{0} }   \ottsym{(}   \ottsym{(}    \mathsf{A_0}   \mathrel{\rightarrow}   \star    \ottsym{)}  \mathrel{\rightarrow}   \star    \ottsym{)}  \mathrel{\rightarrow}   \mathsf{A_0}   
\end{align*}

$ \mathsf{A_0} $ has one constructor whose only argument has type $ \ottsym{(}    \mathsf{A_0}   \mathrel{\rightarrow}   \star    \ottsym{)}  \mathrel{\rightarrow}   \star  $.
Note that we don't need to use its induction principle (\ie recursion),
merely the fact that there's an injection from the latter type to the former,
and so can be seen as a type-theoretic formulation of Reynolds' paradox~\citep{polymorphism-not-set};
this has also been detailed by Coquand~\citep{new-paradox}.

We can define an injection $ \mathsf{ f } $ from $  \mathsf{A_0}   \mathrel{\rightarrow}   \star  $ to $ \mathsf{A_0} $.
Injectivity of both $ \mathsf{MkA_0} $ and $ \mathsf{ f } $ are omitted below;
they are a crucial part of the paradox,
but are orthogonal to what fails to type check.
\begin{align*}
  & \mathsf{ f } :^{ \ottsym{0} }   \ottsym{(}    \mathsf{A_0}   \mathrel{\rightarrow}   \star    \ottsym{)}  \mathrel{\rightarrow}   \mathsf{A_0}    \\
  &   \mathsf{ f }  \;  \mathit{ \mathit{x} }    \coloneqq    \mathsf{MkA_0}   \;  \ottsym{(}    \lambda \mathit{ \mathit{z} } \mathpunct{.}   \mathit{ \mathit{z} }    =   \mathit{ \mathit{x} }    \ottsym{)}  
\end{align*}

Now we are in a position to define a property $ \mathsf{ P } $
similar to regularity from Russell's paradox above,
and an element of $ \mathsf{A_0} $ that simultaneously does and doesn't satisfy $ \mathsf{ P } $.
\begin{align*}
  & \mathsf{ P } :^{ \ottsym{1} }    \mathsf{A_0}   \mathrel{\rightarrow}   \star    \\
  &   \mathsf{ P }  \;  \mathit{ \mathit{x} }    \coloneqq     \mathsf{ NPair }   \;  \ottsym{(}    \mathsf{A_0}   \mathrel{\rightarrow}   \star    \ottsym{)}   \;  \ottsym{(}     \lambda \mathit{ P } \mathpunct{.}   \mathsf{ Pair }    \;  \ottsym{(}     \mathit{ \mathit{x} }   =   \mathit{ f }    \;   \mathit{ P }    \ottsym{)}   \;  \ottsym{(}     \mathit{ P }   \;   \mathit{ \mathit{x} }    \mathrel{\rightarrow}   \bot    \ottsym{)}   \ottsym{)}   \\
  & \mathsf{a_0} :^{ \ottsym{1} }   \mathsf{A_0}   \\
  &  \mathsf{a_0}   \coloneqq    \mathsf{ f }   \;   \mathsf{ P }   
\end{align*}

The details are omitted, but the where the paradox fails to type check
is in trying to construct an element of $  \mathsf{ P }   \;   \mathsf{a_0}  $
using $ \mathsf{ P } $ itself as the first element of the pair.
Its level is $1$, which is too high for the dependent pair,
which asks for a first component at level $0$;
displacing $ \mathsf{ NPair } $ will raise the level of $ \mathsf{ P } $,
which will again make it still too high.

Impredicativity is what normally makes this paradox go through,
disallowing nonstrictly-positive inductives for consistency.
As \lang is predicative, this may permit us to have nonstrictly-positive datatypes consistently;
precedents include Blanqui's Calculus of Algebraic Constructions~\cite[Section~7]{CAC}.

\section{Exponential universe polymorphism} \label{sec:appx:upoly}

\subsection{Coq}

\begin{minted}{coq}
  Set Universe Polymorphism.
  Time Definition T1 : Type := Type -> Type -> Type -> Type -> Type -> Type.
  Time Definition T2 : Type := T1 -> T1 -> T1 -> T1 -> T1 -> T1.
  Time Definition T3 : Type := T2 -> T2 -> T2 -> T2 -> T2 -> T2.
  Time Definition T4 : Type := T3 -> T3 -> T3 -> T3 -> T3 -> T3.
  Time Definition T5 : Type := T4 -> T4 -> T4 -> T4 -> T4 -> T4.
  Time Definition T6 : Type := T5 -> T5 -> T5 -> T5 -> T5 -> T5.
  Time Definition T7 : Type := T6 -> T6 -> T6 -> T6 -> T6 -> T6.
  Time Definition T8 : Type := T7 -> T7 -> T7 -> T7 -> T7 -> T7.
\end{minted}

\subsection{Lean}

\begin{minted}{lean}
  def T1 : Type _ := Type _ → Type _ → Type _ → Type _ → Type _ → Type _
  def T2 : Type _ := T1 → T1 → T1 → T1 → T1 → T1
  def T3 : Type _ := T2 → T2 → T2 → T2 → T2 → T2
  def T4 : Type _ := T3 → T3 → T3 → T3 → T3 → T3
  def T5 : Type _ := T4 → T4 → T4 → T4 → T4 → T4
  def T6 : Type _ := T5 → T5 → T5 → T5 → T5 → T5
  def T7 : Type _ := T6 → T6 → T6 → T6 → T6 → T6
  def T8 : Type _ := T7 → T7 → T7 → T7 → T7 → T7
\end{minted}

\end{document}